# Comparing the Pearson and Spearman Correlation Coefficients Across Distributions and Sample Sizes: A Tutorial Using Simulations and Empirical Data

J. C. F. de Winter[a], S. D. Gosling[b,c], J. Potter[d]

[a]Department of BioMechanical Engineering, Faculty of Mechanical, Maritime and Materials Engineering, Delft University of Technology, The Netherlands, Email: j.c.f.dewinter@tudelft.nl

[b]Department of Psychology, University of Texas at Austin, Austin, TX, USA, Email: samg@austin.utexas.edu

[c]School of Psychological Sciences, University of Melbourne, Parkville, Victoria, Australia

[d]Atof Inc., Cambridge, Massachusetts, jeff@jeffpotter.org

**Abstract**

The Pearson product-moment correlation coefficient ($r_p$) and the Spearman rank correlation coefficient ($r_s$) are widely used in psychological research. We compare $r_p$ and $r_s$ on 3 criteria: variability, bias with respect to the population value, and robustness to an outlier. Using simulations across low ($N = 5$) to high ($N = 1,000$) sample sizes we show that, for normally distributed variables, $r_p$ and $r_s$ have similar expected values but $r_s$ is more variable, especially when the correlation is strong. However, when the variables have high kurtosis, $r_p$ is more variable than $r_s$. Next, we conducted a sampling study of a psychometric dataset featuring symmetrically distributed data with light tails, and of 2 Likert-type survey datasets, 1 with light-tailed and the other with heavy-tailed distributions. Consistent with the simulations, $r_p$ had lower variability than $r_s$ in the psychometric dataset. In the survey datasets with heavy-tailed variables in particular, $r_s$ had lower variability than $r_p$, and often corresponded more accurately to the population Pearson correlation coefficient ($R_p$) than $r_p$ did. The simulations and the sampling studies showed that variability in terms of standard deviations can be reduced by about 20% by choosing $r_s$ instead of $r_p$. In comparison, increasing the sample size by a factor of 2 results in a 41% reduction of the standard deviations of $r_s$ and $r_p$. In conclusion, $r_p$ is suitable for light-tailed distributions, whereas $r_s$ is preferable when variables feature heavy-tailed distributions or when outliers are present, as is often the case in psychological research.

*Keywords:* correlation, outlier, rank transformation, nonparametric versus parametric

# Comparing the Pearson and Spearman Correlation Coefficients Across Distributions and Sample Sizes: A Tutorial Using Simulations and Empirical Data

The Pearson product-moment correlation coefficient ($r_p$; Pearson, 1896) and the Spearman rank correlation coefficient ($r_s$; Spearman, 1904) were developed over a century ago (for a review, see Lovie, 1995). Both coefficients are widely used in psychological research. According to a search of ScienceDirect, of the 18,419 articles published in psychology in 2014, 24.7% reported an effect size measure of some kind. As shown in Table 1, $r_p$ and $r_s$ are particularly popular in sciences involving the analysis of human behavior (social sciences, psychology, neuroscience, medicine). Table 1 further shows that $r_p$ is reported about twice as frequently as $r_s$. Moreover, Table 1 almost certainly underestimates the prevalence of $r_p$, because $r_p$ is the default option in many statistical packages; so when the type of correlation coefficient goes unreported, it is likely to be $r_p$.

Table 1

*Percentage of the papers with abstract published in 2014 that contain a correlation or effect size term, for eight selected subject areas.*

| | 1. Psychology | 2. Neuro-science | 3. Medicine & Dentistry | 4. Social Sciences | 5. Economics, Econome-trics, & Finance | 6. Computer Sciences | 7. Engineering | 8. Chemistry | All eight subject areas |
|---|---|---|---|---|---|---|---|---|---|
| Any of the keywords below | 24.70% | 19.18% | 18.62% | 12.56% | 6.61% | 4.15% | 1.94% | 1.17% | 10.42% |
| ABS({.}) AND ALL("odds ratio" OR "risk ratio" OR "relative risk RR") | 6.80% | 5.60% | 10.37% | 4.21% | 1.76% | 0.46% | 0.35% | 0.08% | 4.88% |
| ABS({.}) AND ALL("Pearson correlation" OR "Pearson product-moment " OR "Pearson r" OR "Pearson's correlation" OR "Pearson's product-moment " OR "Pearson's r") | 9.37% | 7.97% | 4.21% | 4.58% | 2.85% | 1.98% | 0.97% | 0.80% | 3.01% |
| ABS({.}) AND ALL("Spearman rank" OR "Spearman correlation" OR "Spearman rho" OR "Spearman's rank" OR "Spearman's correlation" OR "Spearman's rho" OR "rank-order correlation") | 3.36% | 3.87% | 3.11% | 1.85% | 1.70% | 0.79% | 0.39% | 0.20% | 1.81% |
| ABS({.}) AND ALL("intraclass correlation" OR "intra-class correlation" OR "intraclass r" OR "intra-class r") | 3.24% | 1.66% | 1.63% | 1.32% | 0.20% | 0.19% | 0.11% | 0.03% | 0.85% |
| ABS({.}) AND ALL("Cohen's d" OR "Cohen d" OR "Cohen's effect size") | 4.47% | 2.18% | 0.73% | 1.17% | 0.08% | 0.22% | 0.06% | 0.00% | 0.52% |
| ABS({.}) AND ALL("Cohen's kappa" OR "kappa statistic" OR "Cohen's k" OR "k-statistic") | 1.12% | 0.54% | 0.73% | 0.81% | 0.27% | 0.54% | 0.11% | 0.02% | 0.44% |
| ABS({.}) AND ALL("Kendall tau" OR "Kendall correlation "OR "Kendall's tau" OR "Kendall's correlation") | 0.23% | 0.20% | 0.10% | 0.17% | 0.45% | 0.25% | 0.09% | 0.01% | 0.11% |
| ABS({.}) AND ALL("Hedges's g" OR "Hedges g" OR "Hedges effect size") | 0.58% | 0.23% | 0.10% | 0.06% | 0.01% | 0.03% | 0.01% | 0.00% | 0.06% |
| ABS({.}) AND ALL("Cramer's V" OR "Cramer's phi") | 0.34% | 0.13% | 0.07% | 0.21% | 0.08% | 0.03% | 0.01% | 0.00% | 0.06% |
| ABS({.}) AND ALL("point biserial" OR "point bi-serial") | 0.34% | 0.14% | 0.08% | 0.12% | 0.02% | 0.03% | 0.01% | 0.00% | 0.05% |
| ABS({.}) AND ALL("concordance correlation") | 0.01% | 0.02% | 0.07% | 0.03% | 0.01% | 0.02% | 0.01% | 0.01% | 0.04% |
| ABS({.}) AND ALL("polychoric correlation" OR "tetrachoric correlation" OR "tetrachoric coefficient") | 0.33% | 0.11% | 0.05% | 0.12% | 0.10% | 0.01% | 0.00% | 0.00% | 0.04% |
| ABS({.}) AND ALL("RV coefficient" OR "congruence coefficient" OR "distance correlation" OR "Brownian correlation" OR "Brownian covariance") | 0.09% | 0.13% | 0.02% | 0.03% | 0.04% | 0.06% | 0.03% | 0.04% | 0.04% |
| ABS({.}) AND ALL("Fleiss kappa") | 0.11% | 0.04% | 0.06% | 0.07% | 0.01% | 0.06% | 0.01% | 0.00% | 0.03% |
| ABS({.}) AND ALL("correlation phi" OR "phi correlation" OR "mean square contingency coefficient" OR "Matthews correlation") | 0.08% | 0.04% | 0.03% | 0.03% | 0.02% | 0.14% | 0.03% | 0.04% | 0.03% |
| ABS({.}) AND ALL("correlation ratio" OR "eta correlation") | 0.02% | 0.03% | 0.02% | 0.04% | 0.02% | 0.04% | 0.01% | 0.00% | 0.02% |
| **Total number of publications in 2014** | 18419 | 33758 | 131076 | 32137 | 12261 | 26120 | 64616 | 53604 | 297669* |

*Note.* This table is based on a full-text search of ScienceDirect conducted on October 9, 2015. The horizontal bars within individual cells linearly correspond to the listed percentages. Searching for "correlation coefficient" while excluding all search terms in Table 1 yielded 9,443 papers; in other words, the type of correlation coefficient often goes unreported.

*"All eight subject areas" is not the sum of the eight columns, but the number of articles retrieved when searching in all eight subject areas simultaneously. This number is smaller than the sum of the publications in the eight individual subject areas because some articles are classified in two or more subject areas.

Many more researchers use $r_p$ rather than $r_s$, perhaps because $r_p$ appears to match more closely the linear relationship they aim to estimate. Other reasons why most researchers choose $r_p$ could be because $r_p$ allows for inferences such as calculation of the variance accounted for, or because it is consistent with the methods of available follow-up analyses, such as linear regression (or ANOVA) by least squares or factor analysis by maximum likelihood. Yet another reason for the widespread use of $r_p$ may be that statistical practices are very much determined by what SPSS, R, SAS, MATLAB, and other software manufacturers implement as their default option (Steiger, 2001; 2004). For example, in MATLAB, the command *corr(x,y)* yields the Pearson correlation coefficient between the vectors *x* and *y*. It requires a longer command (*corr(x,y, 'type', 'spearman'*) to calculate the Spearman correlation. Thus, the software may implicitly give the impression that $r_p$ is the preferred option and it also requires more knowledge of the software commands to calculate $r_s$.

### Some Well-Known and Less Well-Known Properties of $r_p$ and $r_s$

The sample Pearson correlation coefficient $r_p$ is defined according to Equation 1. Here, we have first performed a mean centering procedure on the *x* and *y* vectors.

$$r_p = \frac{\sum_{i=1}^{N} x_i y_i}{\sqrt{\sum_{i=1}^{N} x_i^2 \sum_{i=1}^{N} y_i^2}} \qquad (1)$$

The sample Spearman correlation coefficient $r_s$ is calculated in the same manner as $r_p$, except that $r_s$ is calculated after both *x* and *y* have been rank transformed to values between 1 and *N* (Equation 2). When calculating $r_s$, a so-called fractional ranking is used, which means that the mean rank is assigned in case of ties. For example, suppose that the two smallest numbers of *x* are equal, then they will be both ranked as 1.5 (i.e., [1+2]/2). Again, a mean centering is first performed (by subtracting *N*/2+1/2 from each of the two ranked vectors).

$$r_s = \frac{\sum_{i=1}^{N} x_{i,r} y_{i,r}}{\sqrt{\sum_{i=1}^{N} x_{i,r}^2 \sum_{i=1}^{N} y_{i,r}^2}} \qquad (2)$$

Assuming there are no ties, Equation 2 can be rewritten in various formats (Equation 3).

$$r_s = \frac{\sum_{i=1}^{N} x_{i,r}^2 - \frac{1}{2}\sum_{i=1}^{N}\left(x_{i,r} - y_{i,r}\right)^2}{\sum_{i=1}^{N} x_{i,r}^2} = 1 - \frac{\sum_{i=1}^{N}\left(x_{i,r} - y_{i,r}\right)^2}{2\sum_{i=1}^{N} x_{i,r}^2} = 1 - \frac{6\sum_{i=1}^{N}\left(x_{i,r} - y_{i,r}\right)^2}{N\left(N^2-1\right)} = \frac{12}{N\left(N^2-1\right)}\sum_{i=1}^{N} x_{i,r} y_{i,r}$$

(3)

It can be inferred from Equations 1–3 that $r_p$ will be high when the individual points lie close to a straight line, whereas $r_s$ will be high when both vectors have a similar ordinal relationship. As mathematically shown by Yuan and Bentler (2000), the distribution of $r_p$ depends only on the fourth-order moments (or kurtoses) of the two variables, not on their skewness (see also Yuan, Bentler, & Zhang, 2005). After all, $r_p$ is a function of second-order sample moments, and so the variance of $r_p$ is determined by fourth-order moments. The non-parametric measure $r_s$, on the other hand, is relatively robust to heavy-tailed distributions and outliers; all data are transformed to values ranging from 1 to *N*, so the influence function is bounded (Croux & Dehon, 2010). Several of the above characteristics of $r_p$ and $r_s$ are covered in many introductory statistics books and graduate-level psychology programs. Furthermore, a large number of research papers have previously described the differences between $r_p$ and $r_s$, and have confirmed that $r_s$ has attractive robustness properties (e.g., Bishara & Hittner, 2014; Fowler, 1987; Hotelling & Pabst, 1936).

Nonetheless, several characteristics of $r_p$ and $r_s$ may not be well known to researchers, even for the standard scenario of normally distributed variables. The derivation of the probability density function of $r_p$ for bivariate normal variables can be traced back to contributions by Fisher (1915), Sawkins (1944), Hotelling (1951; 1953), and Kenney and Keeping (1951), and was reported more recently by Shieh (2010):

$$f(r_p) = \frac{(N-2)\left(1-R_p{}^2\right)^{\frac{(N-1)}{2}}\left(1-r_p{}^2\right)^{\frac{(N-4)}{2}}}{\sqrt{N}\,(N-2)\,\beta\left(\frac{1}{2},N-\frac{1}{2}\right)\left(1-R_p r_p\right)^{N-\frac{3}{2}}}\,{}_2F_1\left(\frac{1}{2},\frac{1}{2};N-\frac{1}{2};\frac{R_p r_p+1}{2}\right) \qquad (4)$$

Here, $R_p$ is the population Pearson correlation coefficient, β is the beta function, and ${}_2F_1$ is Gauss' hypergeometric function. The hypergeometric function is available in software packages (e.g., *hypergeom([1/2 1/2],N-1/2,(R_p\*r_p+1)/2)* in MATLAB), but can also be readily calculated according to a power series, with Γ being the gamma function:

$${}_2F_1\left(\frac{1}{2},\frac{1}{2};N-\frac{1}{2};\frac{R_p r_p+1}{2}\right) = \sum_{i=0}^{\infty}\left(\frac{\Gamma\left(\frac{1}{2}+i\right)^2\Gamma\left(N-\frac{1}{2}\right)}{\pi\cdot\Gamma\left(N-\frac{1}{2}+i\right)}\frac{\left(\frac{R_p r_p+1}{2}\right)^i}{i!}\right) \qquad (5)$$

Shieh (2010) stated: "It is not well understood that the underlying probability distribution function of $r$ is complicated in form, under the classical assumption that the two variables follow a bivariate normal distribution. The complexity incurs continuous investigation" (p. 906). Figure 1 illustrates the probability density function of $r_p$ for two sample sizes ($N = 5$ and 50) and three population correlation coefficients ($R_p = .2, .4,$ and $.8$). It can be seen that the mode of the distribution is greater than $R_p$ and that the distribution is negatively skewed, with the skew being stronger for higher $R_p$ and for smaller $N$.

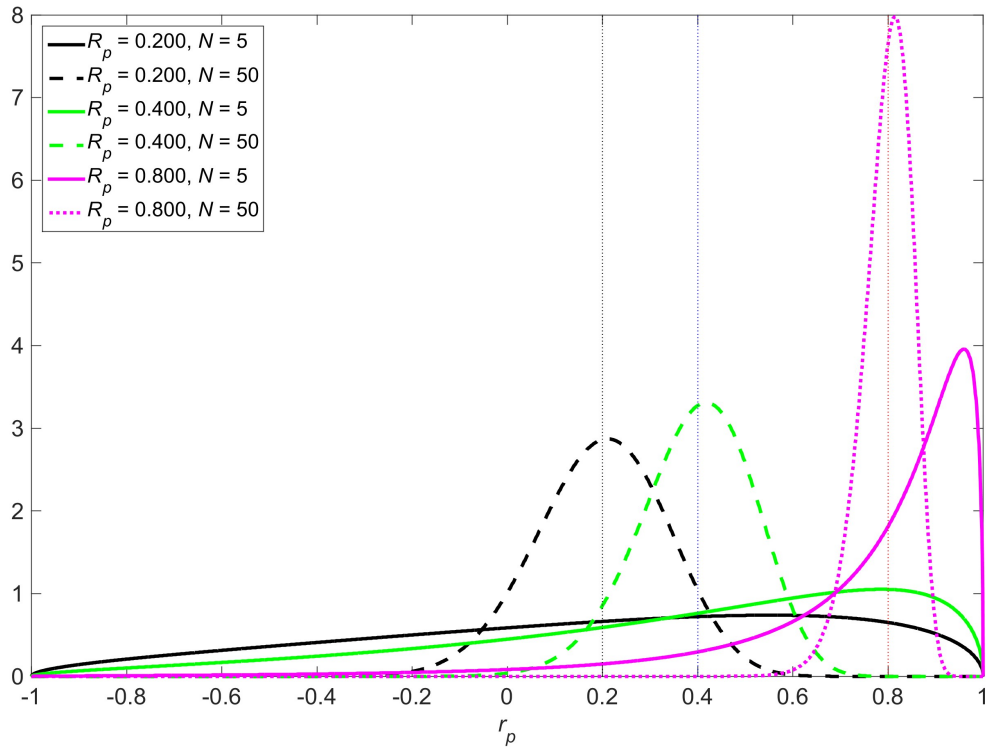

*Figure 1.* Probability density function of the Pearson correlation coefficient ($r_p$) for three levels of the population Pearson correlation coefficient ($R_p = .2$, $R_p = .4$, $R_p = .8$) and two levels of sample size ($N = 5$, $N = 50$). The area under each curve equals 1.

Equation 4 allows one to calculate exact *p*-values and confidence intervals. However, the popular and considerably more straightforward Fisher transformation can also be used in statistical inference (e.g., Fisher, 1921; Fouladi & Steiger, 2008; Hjelm & Norris, 1962; Hotelling, 1953; Winterbottom, 1979). For $r_s$, exact probability density functions are available for small sample sizes, and over the years various approximations (in terms of bias, mean squared error, and relative asymptotic efficiency) of the distribution and its moments have been published (Best & Roberts, 1975; Bonett & Wright, 2000; Croux & Dehon, 2010; David & Mallows, 1961; David, Kendall, & Stuart, 1951; Fieller, Hartley, & Pearson, 1957; Xu, Hou, Hung, & Zou, 2013). Furthermore, several variance-stabilizing transformations have been developed for $r_s$. These transformations, which can be applied in analogous fashion to the Fisher *z*-transformation for $r_p$, may be practical for statistical inference purposes (Bonett & Wright, 2000; Fieller et al., 1957; but see Borkowf, 2002 demonstrating limitations of this concept).

Typically in psychology, investigators undertake research on samples (i.e., a subset of the population) with the aim of estimating the true relationships in the population. It is useful to point out that the expected values of both $r_p$ and $r_s$ are biased estimates of their respective population coefficients $R_p$ and $R_s$ (Ghosh, 1966; Zimmerman, Zumbo, & Williams, 2003). Zimmerman et al. (2003) stated: "It is not widely recognized among researchers that this bias can be as much as .03 or .04 under some realistic conditions" (p. 134). Equation 6 provides the expected value of $r_p$ (Ghosh, 1966), while Equation 7 provides the expected value of $r_s$ (Moran, 1948; Xu et al., 2013; Zimmerman et al., 2003). Both these equations indicate that the population value is underestimated, especially for small *N*. This underestimation is relatively small if $R_p$ is small or moderate. For example, if $R_p = .2$ (corresponding $R_s = .191$, calculated using Equation 9), then $E(r_p)$ and $E(r_s)$ are .177 and .160, respectively at $N = 5$, and .195 and .182 at $N = 20$. The underestimation is more severe for $R_p$ between .3 and .9. If $R_p = .8$ ($R_s = .786$), then $E(r_p)$ and $E(r_s)$ are .754 and .688 at $N = 5$, and .792 and .758 at $N = 20$.

$$E\left(r_p\right)=\frac{2\left(\Gamma\left(\dfrac{N}{2}\right)\right)^2}{(N-1)\left(\Gamma\left(\dfrac{N-1}{2}\right)\right)^2}R_p\cdot\,_2F_1\left(\frac{1}{2},\frac{1}{2};\frac{N+1}{2};R_p^2\right) \tag{6}$$

$$E\left(r_s\right)=\frac{6}{\pi(N+1)}\left(\arcsin\left(R_p\right)+(N-2)\arcsin\left(\frac{R_p}{2}\right)\right) \tag{7}$$

Equation 7 can be rewritten into a form that clarifies how the expected value of $r_s$ relates to the population value of the Spearman coefficient and another well-known rank coefficient, Kendall's tau (Durbin & Stuart, 1951; Hoeffding, 1948).

$$E\left(r_s\right)=\frac{(N-2)R_s+3R_t}{N+1} \tag{8}$$

The Pearson, Spearman, and Kendall correlation coefficients at the population level (i.e., $R_p$, $R_s$, $R_t$) for normally distributed variables can be described by a closed-form expression (e.g., Croux & Dehon, 2010; Pearson, 1907). In other words, for an infinite sample size, the Pearson, Spearman, and Kendall correlation coefficients differ when the two variables are normally distributed (Equations 9, 10, & 11).

$$R_s=\frac{6}{\pi}\arcsin\left(\frac{R_p}{2}\right) \tag{9}$$

$$R_t=\frac{2}{\pi}\arcsin\left(R_p\right) \tag{10}$$

$$R_s=\frac{6}{\pi}\arcsin\left(\frac{\sin\left(\dfrac{1}{2}\pi R_t\right)}{2}\right) \tag{11}$$

The maximum difference between $R_p$ and $R_s$ is .0181 and occurs at $R_p=.594\left(\dfrac{\sqrt{4\pi^2-36}}{\pi}\right)$ and $R_s=.576$

$\left(\dfrac{6}{\pi}\arcsin\left(\dfrac{\sqrt{\pi^2-9}}{\pi}\right)\right)$, see also Guérin, De Oliveira, and Weber (2013). Figure S1 of the supplementary material illustrates the relationships between $R_p$, $R_s$, and $R_t$ (see also Kruskal, 1958).

### Aim of the Present Study

As shown above, the definitions and essential characteristics of $r_p$ and $r_s$ are probably well known. However, $r_p$ and $r_s$ exhibit a variety of interesting features in the case of bivariate normality. Of course, in real-life scenarios, psychologists are likely to encounter non-normal data as well.

In light of the widespread use of correlations in psychology and the predominance of $r_p$ over $r_s$, the goal of this contribution is to review the properties of the $r_p$ versus $r_s$, and to clarify the situations in which $r_p$ or $r_s$ should be preferred. We examine the properties of both coefficients with the aim of providing researchers with empirically derived guidance about which coefficient to use.

We use simulations and analyses of existing datasets to compare $r_p$ with $r_s$ for conditions that are representative of those found in psychological research. We start out by comparing $r_p$ versus $r_s$ for normally distributed variables, which as we indicated above, may have various unfamiliar properties. We aim to depict the characteristics of $r_p$ and $r_s$ in an intuitive, graphical manner. Next, we evaluate $r_p$ versus $r_s$ when the two variables have a non-normal distribution, a situation that is common in psychological research. We also graphically illustrate the strength of $r_s$ when one or more outliers are present. Finally, we provide a demonstration of the differences of $r_p$ versus $r_s$ for typical psychological data. The main contribution of these sampling studies is to explain the relative performance of $r_p$ versus $r_s$ as a function of item/scale characteristics and sample size. In all cases, we compare the two coefficients in terms of variability, bias with respect to the population value, and robustness to an outlier.

### $r_p$ Versus $r_s$ With a Normally Distributed Population

**Normally Distributed Variables in Psychological Research**
The central limit theorem states that the sum of a large number of independent random variables conforms to a normal distribution. Psychologists often aggregate data into constructs, and furthermore, various types of human attributes (such as personality and intelligence) may be seen as the effect of a large number of unobserved random processes. Hence, the central limit theorem can explain why certain psychological variables are *approximately* normally distributed (see Lyon, 2014, for a discussion on the factors that contribute to normality). Intelligence and physical ability are prime examples of human attributes that follow an approximately normal distribution (Burt, 1957; Plomin & Deary, 2015). The normal distribution occurs empirically regardless of whether the attribute is measured on an ordinal scale (e.g., a paper and pencil intelligence test) or on a ratio scale (e.g., intelligence defined chronometrically; Jensen, 2006). Let us therefore first evaluate how $r_p$ and $r_s$ behave when the two variables are normally distributed.

**Selected Population Correlation Coefficients**
To describe the behavior of $r_p$ and $r_s$ for bivariate normal variables and finite sample sizes, we undertook a simulation study. To ensure that the ranges of coefficient sizes were representative of those potentially encountered in psychological research, we consulted the literature. In published research, correlations among psychometric test scores, and correlations between psychological assessment scores and performance criteria, generally range between 0 and .5 (cf. Jensen, 2006; Meyer et al., 2001; Tett, Jackson, & Rothstein, 1991). One review of 322 meta-analyses showed that the absolute correlation coefficients in social psychology average at .21, with 95% of the coefficients between 0 and .5, and the remaining 5% between .5 and .8 (Richard, Bond, & Stokes-Zoota, 2003). Only variables that are conceptually similar to one another, such as intelligence test scores and scholastic performance, will correlate as highly as .8 (Deary, Strand, Smith, & Fernandes, 2007; Frey & Detterman, 2004). In short, population correlations between 0 and .8 reflect the range found in virtually all psychological/behavioral research. Therefore, simulation studies were performed with population Pearson correlation coefficients that were zero ($R_p = 0$), moderate ($R_p = .2$), strong ($R_p = .4$), and very strong ($R_p = .8$). The corresponding population Spearman correlation coefficients ($R_s$) were calculated according to Equation 9.

**Selected Sample Sizes**
Sample sizes used by psychologists are known to vary widely. One analysis of hundreds of articles (Marszalek, Barber, Kohlhart, & Holmes, 2011) showed that in the *Journal of Experimental Psychology* in the year 2006, the median total sample size was 18 ($Q_1 = 10$, $Q_3 = 32$), whereas in the *Journal of Applied Psychology*, the mean sample size was 148 ($Q_1 = 45$, $Q_3 = 269$). Fraley and Vazire (2014) showed that the median sample size in five high-impact psychological journals in the years 2006–2010 ranged between 73 ($Q_1 = 41$, $Q_3 = 143$) for *Psychological Science* and 178 ($Q_1 = 100$, $Q_3 = 344$) for the *Journal of Personality* (we calculated the interquartile ranges from the supplementary material of Fraley & Vazire, 2014). Here we note that personality psychology is more likely than experimental psychology to use correlation coefficients (e.g., Cronbach, 1957; Tracy, Robins, Sherman, 2009), and so a sample size of about 200 is regarded as typical for correlational analyses. This sample size is in line with a recent simulation study that investigated at which sample size correlations stabilize, and which concluded that "there are few occasions in which it may be justifiable to go below $n = 150$ and for typical research scenarios reasonable trade-offs between accuracy and confidence start to be achieved when *n* approaches 250" (Schönbrodt & Perugini, 2013, p. 611).

To cover the range of sample sizes found in psychological research, we used 25 sample sizes (*N*s) logarithmically spaced between 5 and 1,000. To generate stable estimates of $r_p$ and $r_s$, for each sample size, 100,000 samples of variable 1 (hereafter called *x*) and variable 2 (hereafter called *y*) were drawn, and $r_p$ and $r_s$ were calculated for each of the 100,000 samples.

## Results of the Simulations

The simulation results for $R_p = .2$ are shown in Figure 2. The mean $r_s$ is slightly lower than the mean $r_p$, for all sample sizes. For small sample sizes, the mean $r_p$ and mean $r_s$ are both slight underestimates of their respective population values $R_p$ and $R_s$ (see also Equations 6 and 7). Figure 2 also shows how the absolute variability decreases with sample size for both $r_p$ and $r_s$. However, $r_s$ has a slightly higher variability, with the standard deviation of $r_s$ being about 0.7% greater than the standard deviation of $r_p$, for each tested sample size. Similarly, the root mean squared error (RMSE) of $r_s$ with respect to $R_s$ is 0.7% greater than the RMSE of $r_p$ with respect to $R_p$.

Note that $r_s$ can take on only a distinct number of values, rapidly increasing with increasing *N* (Sloane, 2003; sequence A126972). For example, for *N* = 5, $r_s$ can be only 1 of 21 different values (−1, −.9, −.8, ..., .8, .9, 1; see Figure S2 for an illustration of the distribution of $r_p$ and $r_s$ at *N* = 5). The supplementary material (Figures S3, S4, and S5) includes the distributions of $r_p$ and $r_s$ for $R_p = 0$, $R_p = .4$, and $R_p = .8$. For $R_p = 0$, $r_p$ and $r_s$ behave almost identically. For $R_p = .4$, the standard deviation of $r_s$ is 3 to 4% higher than the standard deviation of $r_p$, and for $R_p = .8$, the standard deviation of $r_s$ is as much as 18% higher than the standard deviation of $r_p$. The smaller variability of $r_p$ compared to $r_s$ is consistent with previous research (Bonett & Wright, 2000; Croux & Dehon, 2010; Fieller et al., 1957) and suggests that when both population variables are known to have approximately normal distributions, $r_p$ should be used instead of $r_s$, especially when the correlation is thought to be strong.

### $r_p$ Versus $r_s$ With a Non-Normally Distributed Population

## Non-Normally Distributed Variables in Psychological Research

It frequently happens that psychological measurements feature a non-normal distribution. For example, it is known that psychiatric and other types of disorders follow a skewed distribution among individuals (Delucchi & Bostrom, 2004; Keats & Lord, 1962; McGrath, Saha, Welham, El Saadi, MacCauley, & Chant, 2004). Yet in other cases, measurement scales may be limited by artefacts such as ceiling and floor effects (Van den Oord, Pickles, & Waldman, 2003). One analysis of 693 distributions of cognitive measures and other psychological variables with sample sizes ranging from 10 to 30 showed that 39.9% of the distributions were considered as slightly non-normal, 34.5% as moderately non-normal, 10.4% as highly non-normal, and a further 9.6% as extremely non-normal (Blanca, Arnau, López-Montiel, Bono, & Bendayan, 2013). Another analysis of 440 large-sample distributions of achievement and psychometric data classified 31% of the distributions as extremely asymmetric, and 49% as having at least one extremely heavy tail (Micceri, 1989).

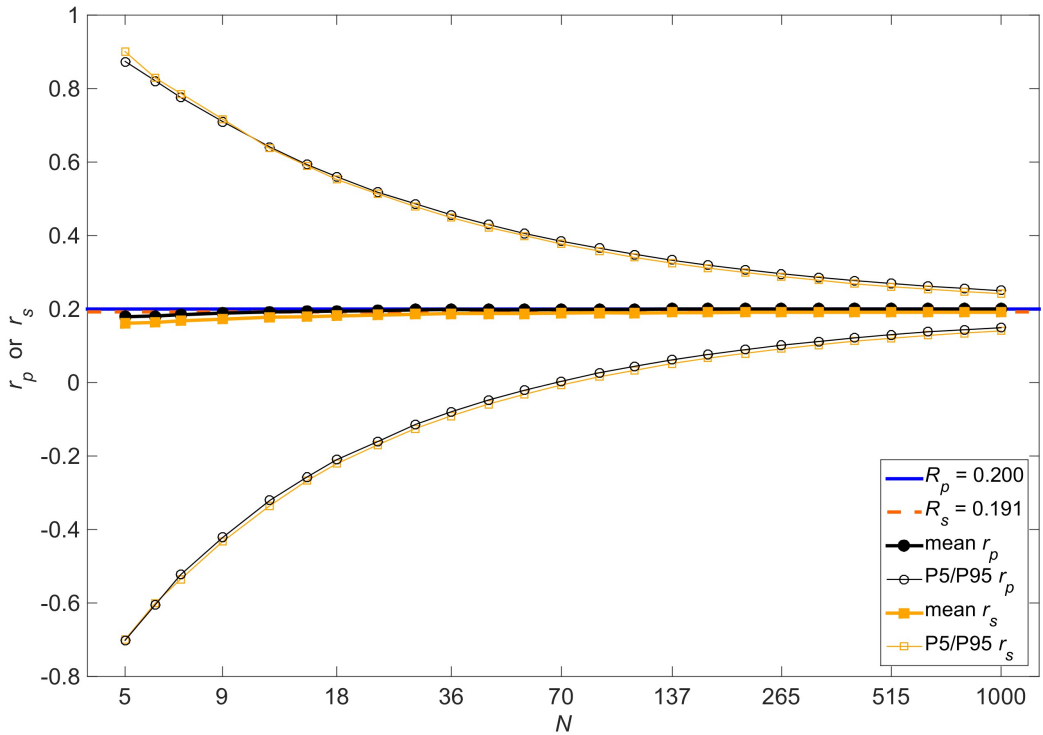

*Figure 2.* Simulation results for normally distributed variables having a population Pearson correlation coefficient of .2 ($R_p = .2$). The population Spearman correlation coefficient ($R_s$) was calculated according to Equation 9. The figure shows the mean, 5th percentile (P5), and 95th percentile (P95) of the Pearson correlation coefficient ($r_p$) and the Spearman correlation coefficient ($r_s$) as a function of sample size ($N$).

### Selected Kurtosis of the Marginal Distributions

In light of these kinds of observations, we explored the behavior of $r_p$ and $r_s$ for two correlated variables having leptokurtic distributions, meaning that kurtosis was greater than would be expected from a normal distribution (see Figure 3 for illustration, and DeCarlo, 1997, for an explanation of kurtosis). The variables $x$ and $y$ were approximately exponentially distributed (hence, skewness = 2 & kurtosis = 9) and strongly correlated ($R_p = .4$). We used a fifth-order polynomial transformation method for generating the correlated non-normally distributed variables (Headrick, 2002). Because $R_p$ and $R_s$ could not be determined exactly, we defined these parameters by calculating the correlation coefficients for a very large sample size ($N = 10^7$).

### Results of the Simulations

Figure 4 shows the distributions of $r_p$ and $r_s$ for the same range of sample sizes as those used to create Figure 2. It can be seen that the expected values of $r_p$ and $r_s$ are about the same and unbiased with respect to their respective population values, but $r_p$ is more variable than $r_s$. Specifically, the standard deviation of $r_p$ is 13.5%, 26.0%, and 27.3% greater than the standard deviation of $r_s$, for $N = 18$, $N = 213$, and $N = 1,000$, respectively. Similarly, the RMSE of $r_p$ with respect to $R_p$ is 13.0%, 25.9%, and 27.3% greater than the RMSE of $r_s$ with respect to $R_s$, for $N = 18$, $N = 213$, and $N = 1,000$, respectively.

### Additional Simulation Results With Other Kurtosis and $R_p$

If the two variables have greater kurtosis than exponentially distributed variables, then $r_p$ is likely to be even more variable (see Figure S6 of the supplementary material). Also note that the size of the correlation coefficient is an important determinant of the behavior of $r_p$ and $r_s$. For example, when choosing $R_p = .2$ instead of $R_p = .4$, the standard deviation of $r_p$ is only 8.0%, 14.5%, and 15.5% greater than the standard deviation of $r_s$, for $N = 18$, $N = 213$, and $N = 1,000$, respectively. However, for $R_p = .8$, the standard deviation of $r_p$ is 13.5%, 36.0%, and 38.9% greater than the standard deviation of $r_s$, for $N = 18$, $N = 213$, and $N = 1,000$, respectively (see Figures S10–S13).

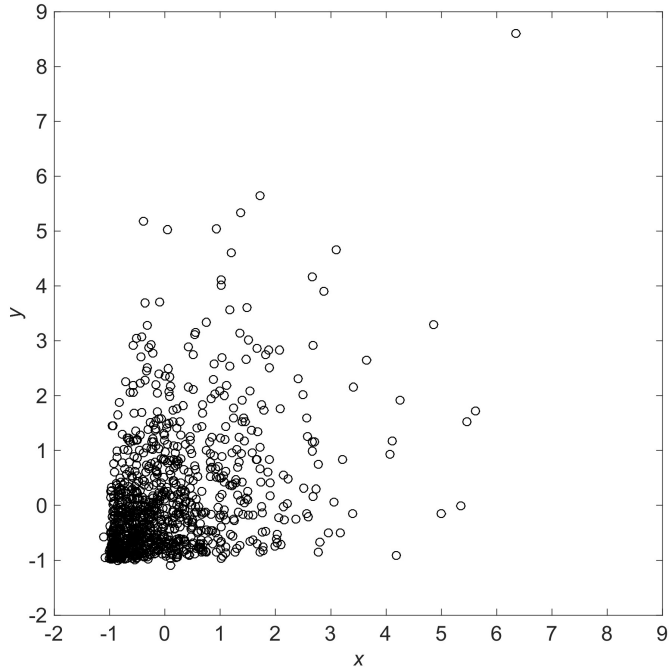

*Figure 3.* Depiction (using $N = 1,000$) of two correlated variables having an exponential distribution with population Pearson correlation coefficient ($R_p$) of .4. $R_p$ was obtained by calculating $r_p$ for a sample of $N = 10^7$ pairs.

In summary, our simulations showed that when the two variables have leptokurtic distributions, $r_p$ is likely to be more variable than $r_s$. These observations are consistent with theory showing that the standard deviation of $r_p$ is proportional to the kurtosis of the variables (Yuan & Bentler, 2000). Moreover, our results are in line with several simulation studies which demonstrated lower variability of $r_s$ compared to $r_p$ for (severely) non-normal distributions (Bishara & Hittner, 2014; Chok, 2010; Kowalski, 1972). Obviously, our set of simulations provide only a snapshot of the constellation of the bivariate relationships that may occur in psychological research. Furthermore, note that when the two variables are mesokurtic or platykurtic (i.e., kurtosis ≤ 3), $r_p$ will tend to be more stable than $r_s$.

### $r_p$ Versus $r_s$ When There Are Outliers

It has been well documented that the Pearson correlation coefficient is sensitive to outliers (e.g., Chok, 2010; Croux & Dehon, 2010). Formal treatments of so-called "influence functions" or "expected resistance" of $r_p$ and $r_s$ can be found in Blair and Lawson (1982), Zayed and Quade (1997), and Croux and Dehon (2010). Herein, we graphically and numerically illustrate how $r_p$ and $r_s$ respond to adding a spurious data point in conditions that are likely to occur in psychological research.

Although sample sizes in psychological research vary widely, we used $N = 200$ because this is in line with typical sample sizes used in applied and personality psychology (Fraley & Vazire, 2014; Marszalek et al., 2011). A sample ($N = 200$) was drawn from two standard normal distributions having a moderate interrelationship in the population ($R_p = .2$). Next, one data point was added so that $N = 201$. The value of the spurious data point was systematically varied from −5 to 5 with a resolution of 0.05 for the two variables, $x$ and $y$. Accordingly, 40,401 (i.e., 201 x 201) $r_p$s and 40,401 $r_s$s were determined.

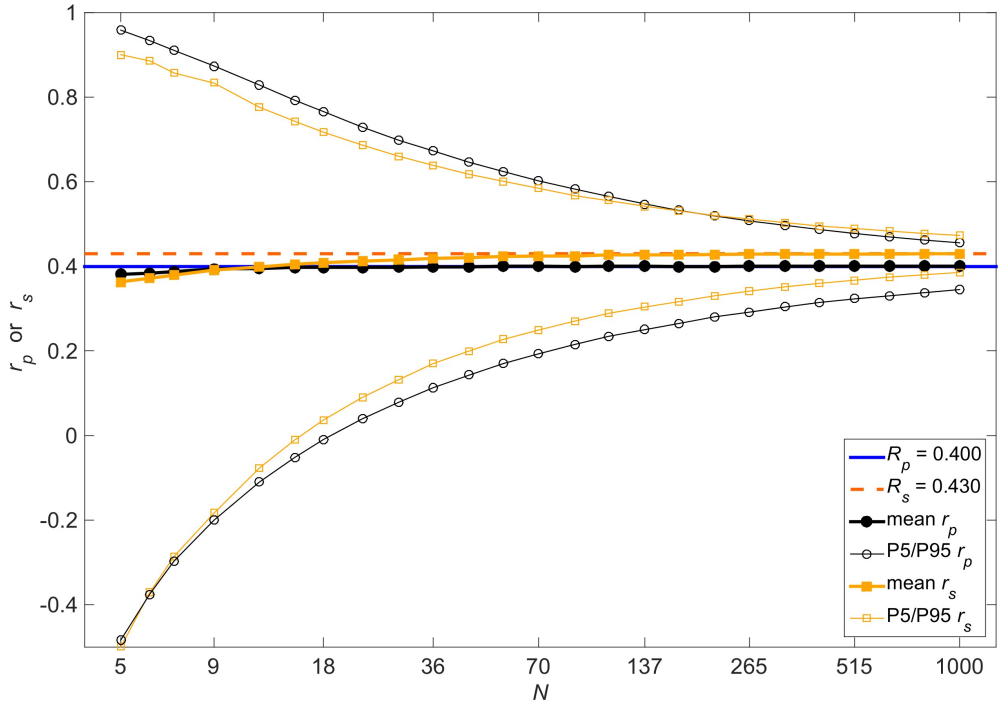

*Figure 4*. Simulation results for two correlated variables having an exponential distribution (see Figure 3 for a large-sample illustration of the distribution). The figure shows the mean, 5th percentile (P5), and 95th percentile (P95) of the Pearson correlation coefficient ($r_p$) and the Spearman correlation coefficient ($r_s$) as a function of sample size ($N$). The population coefficients $R_p$ and $R_s$ were obtained by calculating $r_p$ and $r_s$, respectively, for a sample of $N = 10^7$.

Figure 5 illustrates the influence of the added (201st) data point on the obtained $r_p$ and $r_s$, respectively. It can be seen that $r_p$ is sensitive to this data point. Specifically, $r_p$ equaled .231 without the data point, and has values between .100 (at $x = −5$, $y = 5$) and .312 (at $x = 5$, $y = 5$) by including it, with 19% of the $r_p$s differing by more than .05 from the original $r_p$ of .231. In contrast, $r_s$ is robust: $r_s$ equals .222 without the extra data point, and adding it results in $r_s$ values between .204 and .233. $r_s$ is robust to outliers because the data in $x$ and $y$ are transformed to integers between 1 and $N$. This means it is impossible for very low or very high values in $x$ or $y$ to have a large effect on $r_s$.

Of course, in most real data there may be more than one outlier. Suppose, for example, that one outlier is located at $x = 5$ and $y = 5$, then adding a second outlier at all possible positions between −5 and 5 results in an $r_p$ ranging between .186 and .377 ($N = 202$), with 77% of the $r_p$s differing by more than .05 from the original $r_p$ of .231. Now suppose that the first outlier is at $x = 5$ and $y = −5$, then adding the second outlier results in an $r_p$ between −.003 and .191. Again, $r_s$ is robust, and always between .186 and .245 when two outliers were present. So, having more than one outlier can create even more problems for $r_p$, as the second outlier does not alleviate the distortive effect of the first outlier.

### Five Demonstrations Using Empirical Data

The simulations above are indicative of the differences between $r_p$ and $r_s$ for normally and non-normally distributed variables. However, the simulations do not necessarily reflect situations encountered by empiricists. To test $r_p$ versus $r_s$ on data likely to be found in psychological studies, we undertook a sampling study using empirical data.

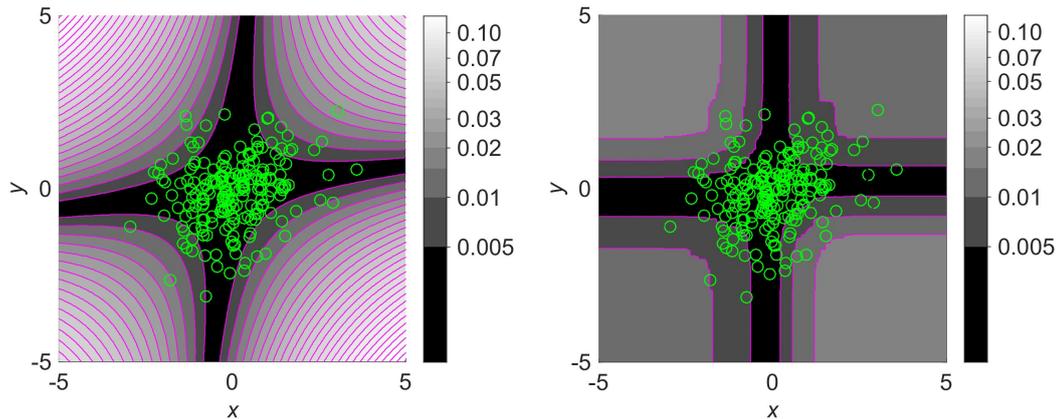

*Figure 5.* Simulation results demonstrating the influence of a spurious data point at location $(x, y)$ on the Pearson correlation coefficient (left figure) and on the Spearman correlation coefficient (right figure). The circles represent a sample ($N = 200$) drawn from two standard normal distributions with population Pearson correlation coefficient ($R_p$) = .2. The sample Pearson correlation coefficient ($r_p$) = .231. The sample Spearman correlation coefficient ($r_s$) = .222. The grayscale background represents the absolute deviation from $r_p$ (left figure) and the absolute deviation from $r_s$ (right figure), after adding one data point so that $N = 201$. Isolines are drawn at every 0.005 increment. The vertical bars next to each figure signify the numeric values corresponding to a particular level of grayness. The value of the data point was systematically varied from $-5$ and $+5$ with a resolution of 0.05 for the two variables, $x$ and $y$.

**Selected Datasets**

Three large datasets were used: a psychometric test battery (Armed Services Vocational Aptitude Battery; ASVAB), and two survey-based datasets: 5-point Likert-scale data from the Big Five Inventory (BFI) and 6-point-scale data from the Driver Behaviour Questionnaire (DBQ). The ASVAB, BFI, and DBQ datasets were all large ($N = 11,878$, $N = 1,895,753$, and $N = 9,077$, respectively), and were therefore used as populations from which we could draw samples to calculate sample correlation coefficients. The ASVAB consists of 10 very strongly intercorrelated test results, each symmetrically distributed with light tails (Table 2). Recall that the simulation results above showed that $r_p$ is less variable than $r_s$ for normally distributed variables that are strongly correlated, so we expected the ASVAB sampling results to reflect these findings. The primary difference between the BFI and DBQ is that the BFI items have low kurtosis because the means of all 44 items are close to the middle option on the five-point scale (Table 2). In contrast, the DBQ items are leptokurtic, with the majority of participants reporting that they "never" make a certain error or violation in traffic (see also Mattsson, 2012). In light of the above simulation results, we expected $r_s$ to outperform $r_p$ for the DBQ dataset, and to a lesser extent for the BFI dataset.

**Sampling Study 1: ASVAB.** The ASVAB dataset is a psychometric dataset consisting of 11,878 subjects who, in the framework of the National Longitudinal Survey of Youth 1979, had taken a test battery (Bureau of Labor Statistics, 2002; Frey & Detterman, 2004; Maier & Sims, 1986; Morgan, 1983). The population included 5,951 men and 5,927 women. The mean age of the subjects was 18.8 years ($SD = 2.3$). The ASVAB consists of 10 tests (general science [25 items], arithmetic reasoning [30 items], word knowledge [35 items], paragraph comprehension [15 items], numerical operations [50 items], coding speed [84 items], auto and shop information [25 items], mathematics knowledge [25 items], mechanical comprehension [25 items], and electronics information [10 items]). The Pearson correlation matrix among the 10 variables contained 45 (=10*(10−1)/2) unique elements. The maximum $R_p$ was .825, occurring between "general science" and "word knowledge" (corresponding $R_s$ = .834). The distribution of the variables was symmetric and platykurtic, that is, having somewhat lighter tails than would be expected from a normal distribution (Table 2).

**Sampling Study 2: BFI items.** The BFI is a 44-item personality questionnaire answered on a Likert scale from 1 = *disagree strongly* to 5 = *agree strongly*. The BFI data ($N = 3,093,144$) were obtained via noncommercial, advertisement-free Internet websites between 1999 and 2013 as part of the Gosling-Potter Internet Personality Project (e.g., Bleidorn et al., 2013; Gosling, Vazire, Srivastava, & John, 2004; Obschonka, Schmitt-Rodermund,

Silbereisen, Gosling, & Potter, 2013; Rentfrow et al., 2013; Srivastava, John, Gosling, & Potter, 2003). Only participants who filled in the English version of the inventory, who answered all items without giving identical answers to all 44 items, and who were between 18 to 98 years were included, leaving a dataset of 1,895,753 respondents. The mean age of the respondents was 28.2 (median = 25.0, $SD$ = 10.4). The population included 921,670 women and 651,914 men, and the sex was unknown for a further 322,169 respondents. The average mean response across the 44 items was 3.45 ($SD$ = 0.47), with a minimum mean of 2.48 for the item "is depressed blue" and a maximum mean of 4.33 for the item "is a reliable worker." The BFI correlation matrix contained 946 (= 44*(44−1)/2) unique off-diagonal elements. The maximum $R_p$ was .597, occurring between "is talkative" and "is outgoing, sociable" (corresponding $R_s$ = .595). The variables were symmetric with low kurtosis (Table 2).

**Sampling Study 3: BFI scales.** Psychological researchers often conduct their analysis at the scale level instead of the item level, so we also carried out the sampling study based on the five BFI scales. The following five sum scores were calculated: agreeableness (9 items), conscientiousness (9 items), extraversion (8 items), openness (10 items), and neuroticism (8 items). The 10 off-diagonal $R_p$s ranged between −.32 (for agreeableness vs. neuroticism; corresponding $R_s$ = −.30) and .28 (for agreeableness vs. conscientiousness; corresponding $R_s$ = .28). Table 2 shows that the five scales were fairly symmetric with low kurtosis.

Table 2
*Means, standard deviations, minima, and maxima of absolute population correlation coefficients, and of population skewness and population kurtosis of the items/scales.*

| Measure | | ASVAB, 45 correlations | BFI items, 946 correlations | BFI scales, 10 correlations | DBQ items, 561 correlations | DBQ scales, 10 correlations |
|---|---|---|---|---|---|---|
| $\mathbf{|R_p|}$ | Mean | .6273 | .1206 | .1778 | .1713 | .4197 |
| | $SD$ | .1205 | .1146 | .0985 | .0790 | .1363 |
| | Min | .3317 | .0002 | .0283 | .0003 | .1511 |
| | Max | .8247 | .5973 | .3158 | .5106 | .5805 |
| | | | | | | |
| $\mathbf{|R_s|}$ | Mean | .6281 | .1223 | .1690 | .1622 | .4157 |
| | $SD$ | .1207 | .1148 | .0935 | .0748 | .1155 |
| | Min | .3362 | .0001 | .0309 | .0024 | .1742 |
| | Max | .8336 | .6029 | .3035 | .4747 | .5362 |
| | | | | | | |
| | | ASVAB, 10 tests | BFI, 44 items | BFI, 5 scales | DBQ, 34 items | DBQ, 5 scales |
| **Skewness** | Mean | −0.02 | −0.37 | −0.21 | 2.19 | 1.65 |
| | $SD$ | 0.40 | 0.42 | 0.10 | 1.44 | 0.72 |
| | Min | −0.59 | −1.33 | −0.30 | 0.50 | 0.83 |
| | Max | 0.50 | 0.42 | −0.06 | 6.42 | 2.46 |
| | | | | | | |
| **Kurtosis** | Mean | 2.32 | 2.54 | 2.89 | 11.96 | 8.89 |
| | $SD$ | 0.18 | 0.67 | 0.15 | 13.90 | 5.17 |
| | Min | 2.03 | 1.80 | 2.74 | 3.16 | 4.03 |
| | Max | 2.73 | 4.73 | 3.08 | 60.89 | 16.61 |

*Note.* $R_p$ = population Pearson correlation coefficient, $R_s$ = population Spearman correlation coefficient, ASVAB = Armed Services Vocational Aptitude Battery, BFI = Big Five Inventory, DBQ = Driver Behaviour Questionnaire. Skewness was defined as the third central moment divided by the cube of the standard deviation. Kurtosis was defined as the fourth central moment divided by the fourth power of the standard deviation. Kurtosis of a normal distribution = 3. $R_p$ and $R_s$ were defined as the correlation coefficients for the total sample (i.e., $N$ = 11,878 for the ASVAB, $N$ = 1,895,753 for the BFI, and $N$ = 9,077 for the DBQ). The population skewness and population kurtosis have a strong correlation (ASVAB: $r_s$ between skewness and kurtosis = −.50 [$N$ = 10 items]), BFI items: $r_s$ = −.83 [$N$ = 44 items], BFI scales: $r_s$ = −.70 [$N$ = 5 scales], DBQ items: $r_s$ = .99 [$N$ = 34 items], DBQ scales: $r_s$ = 1.00 [$N$ = 5 scales]).

**Sampling Study 4: DBQ items.** The DBQ dataset consisted of 9,077 respondents who, as part of a cohort study of learner and new drivers, had responded to the query "when driving, how often do you do each of the following?" with respect to 34 items (Transport Research Laboratory, 2008; Wells, Tong, Sexton, Grayson, & Jones, 2008). The responses ranged from 1 = *never* to 6 = *nearly all the time.* The mean age of the respondents was 22.6 years (median = 18.7; $SD$ = 8.1). The population consisted of 5,754 women and 3,323 men. The average mean response across the 34 items was 1.46 ($SD$ = 0.26), with a minimum mean of 1.05 and a maximum mean of 2.06. The correlation matrix contained 561 (= 34*(34−1)/2) unique off-diagonal elements. The maximum $R_p$ was .511 (between "Disregard the speed limit on a motorway" and "Disregard the speed limit on a residential road") with a corresponding $R_s$ of .475. Items were highly skewed and leptokurtic (Table 2).

**Sampling Study 5: DBQ scales.** The DBQ analysis was repeated at the scale level. The following five sum scales were calculated (as in Wells et al., 2008): violations (6 items), errors (8 items), aggressive violations (6 items), inexperience errors (7 items), and slips (7 items). The 10 off-diagonal $R_p$s ranged between .151 (between aggressive violations and inexperience errors; corresponding $R_s$ = .174) and .581 (between violations and aggressive violations; corresponding $R_s$ = .536). As with the DBQ items, the DBQ scales had high kurtosis, but the scale data were more strongly intercorrelated than the item data (Table 2).

**Sampling Methods**
For each of the five datasets (i.e., ASVAB, BFI items, BFI scales, DBQ items, and DBQ scales), 50,000 random sample of $N$ = 200 were drawn with replacement. For each drawn sample, the Pearson and Spearman correlation matrices were calculated. Next, for each element of the correlation matrices, we calculated the absolute of the mean and the standard deviation across the 50,000 samples. To assess how accurately the sample correlation coefficients corresponded to the population values, we calculated the mean absolute difference of each $r_p$ and $r_s$ with respect to the population values ($R_p$ and $R_s$). $R_p$ and $R_s$ were defined as the correlation coefficients for the full population ($N$ = 11,878 for the ASVAB, $N$ = 1,895,753 for the BFI, and $N$ = 9,077 for the DBQ).

**Results of the Five Sampling Studies**
A numerical comparison between the performance of $r_p$ and $r_s$ is provided in Table 3. It can be seen that for the ASVAB data, $r_p$ gives the same average values as $r_s$, with about 6% lower variability (i.e., lower $SD$). For the BFI and DBQ data, the opposite results were found: the mean absolute difference between $r_s$ and $R_s$ is smaller than the mean absolute difference between $r_p$ and $R_p$. In other words, Spearman correlation coefficients are closer to their population value than are Pearson correlation coefficients. Furthermore, for the DBQ data in particular, the mean absolute difference between $r_s$ and $R_p$ is smaller than the mean absolute difference between $r_p$ and $R_p$. That is, $r_s$ even outperformed $r_p$ in recovering $r_p$'s own population value.

Table 3 further shows that the superior performance of $r_s$ is evident for the DBQ dataset (featuring kurtosis > 3 for all items) and is less evident for the BFI dataset (featuring average kurtosis < 3). $r_p$ on average has 2% higher variability (i.e., higher $SD$) than $r_s$ for the BFI items, 4% higher variability for the BFI scales, 18% higher variability for the DBQ items, and 24% higher variability for the DBQ scales.

Table 3

*Means and standard deviations of sample correlation coefficients, and mean absolute difference between sample correlation coefficients and population correlation coefficients (N = 200).*

| Measure | ASVAB Mean across 45 correlations | BFI items Mean across 946 correlations | BFI scales Mean across 10 correlations | DBQ items Mean across 561 correlations | DBQ scales Mean across 10 correlations |
|---|---|---|---|---|---|
| \|Mean $r_p$\| | .6269 | .1205 | .1772 | .1694 | .4178 |
| \|Mean $r_s$\| | .6258 | .1221 | .1683 | .1616 | .4144 |
| \|Mean $r_p$\|−\|$R_p$\| | −0.0005 | −0.0001 | −0.0005 | −0.0019 | −0.0019 |
| \|Mean $r_s$\|−\|$R_s$\| | −0.0022 | −0.0002 | −0.0008 | −0.0006 | −0.0013 |
| $SD\ r_p$ | .0411 | .0732 | .0741 | .0872 | .0750 |
| $SD\ r_s$ | .0436 | .0715 | .0714 | .0742 | .0605 |
| Mean \|$r_p$−$R_p$\| | .0327 | .0585 | .0592 | .0697 | .0596 |
| Mean \|$r_p$−$R_s$\| | .0352 | .0587 | .0602 | .0701 | .0628 |
| Mean \|$r_s$−$R_p$\| | .0371 | .0574 | .0579 | .0606 | .0522 |
| Mean \|$r_s$−$R_s$\| | .0347 | .0571 | .0570 | .0593 | .0483 |

*Note.* $R_p$ = population Pearson correlation coefficient; $R_s$ = population Spearman correlation coefficient; ASVAB = Armed Services Vocational Aptitude Battery; BFI = Big Five Inventory; DBQ = Driver Behaviour Questionnaire. Skewness was defined as the third central moment divided by the cube of the standard deviation. Kurtosis was defined as the fourth central moment divided by the fourth power of the standard deviation. Kurtosis of a normal distribution = 3. $R_p$ and $R_s$ were defined as the correlation coefficients for the total sample (i.e., $N$ = 11,878 for the ASVAB, $N$ = 1,895,753 for the BFI, & $N$ = 9,077 for the DBQ).

The mean absolute difference of $r_p$ (and to a lesser extent of $r_s$) with respect to the population value is particularly large for pairs of DBQ items that have distributions with high kurtosis (see Figure S7 of the supplementary material). The distributions of $r_p$ and $r_s$ for the two DBQ items having the highest kurtosis (60.9 and 57.2, respectively) are illustrated in Figure 6. It can be seen that for this selected pair of variables, $r_p$ was considerably more variable than $r_s$, with the standard deviation at $N$ = 1,000 being .071 for $r_p$ and .049 for $r_s$. Figure 7 illustrates the variability of $r_p$ and $r_s$ as a function of $R_p$ for each of the five sampling studies. It can be seen that $r_s$ is considerably less variable than $r_p$, especially for the BFI scales, DBQ items, and DBQ scales.

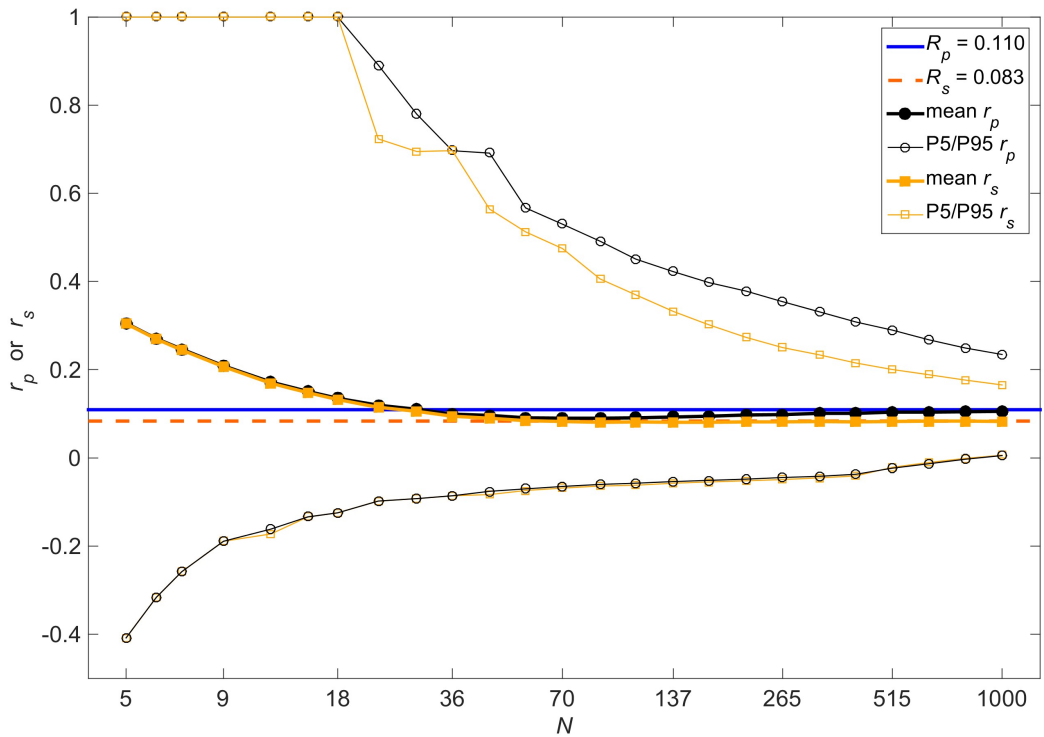

*Figure 6.* Sampling results for the two variables of the Driver Behaviour Questionnaire (DBQ) having the highest kurtosis of the 34 items (population kurtosis = 60.9 and 57.2, respectively; population skewness = 6.42 and 6.05, respectively). The figure shows the mean, 5th percentile (P5), and 95th percentile (P95) of the Pearson correlation coefficient ($r_p$) and the Spearman correlation coefficient ($r_s$) as a function of sample size ($N$). The population coefficients $R_p$ and $R_s$ were defined as the correlation coefficients for the total sample ($N$ = 9,077). The results were based on 50,000 samples. Note that 8,272 of 9,077 respondents answered "never" to both items, and hence the correlation coefficient could often not be calculated when the sample size was small. The sampling was repeated when the correlation coefficient could not be calculated.

### Additional Simulations With *N* = 25 and *N* = 1,000

The results in Table 3 and Figure 7 were based on a sample size of 200. To test whether the results depend on sample size, the simulations were repeated for *N* = 25 and *N* = 1,000 (Tables S1 & S2 in the supplementary material). For *N* = 25, the variabilities of $r_p$ and $r_s$ are obviously higher than for *N* = 200, but the pattern of differences between $r_p$ and $r_s$ is the same. For *N* = 1,000, the variabilities of $r_p$ and $r_s$ are considerably lower than for *N* = 200, but again the pattern of differences is the same, with $r_s$ having a lower standard deviation than $r_p$ for the BFI and DBQ datasets. For *N* = 1,000 it is less likely that the mean absolute difference between $r_s$ and $R_p$ is smaller than the mean absolute difference between $r_p$ and $R_p$, because at such high sample size, the correlation coefficients $r_p$ and $r_s$ are close to their own respective population values.

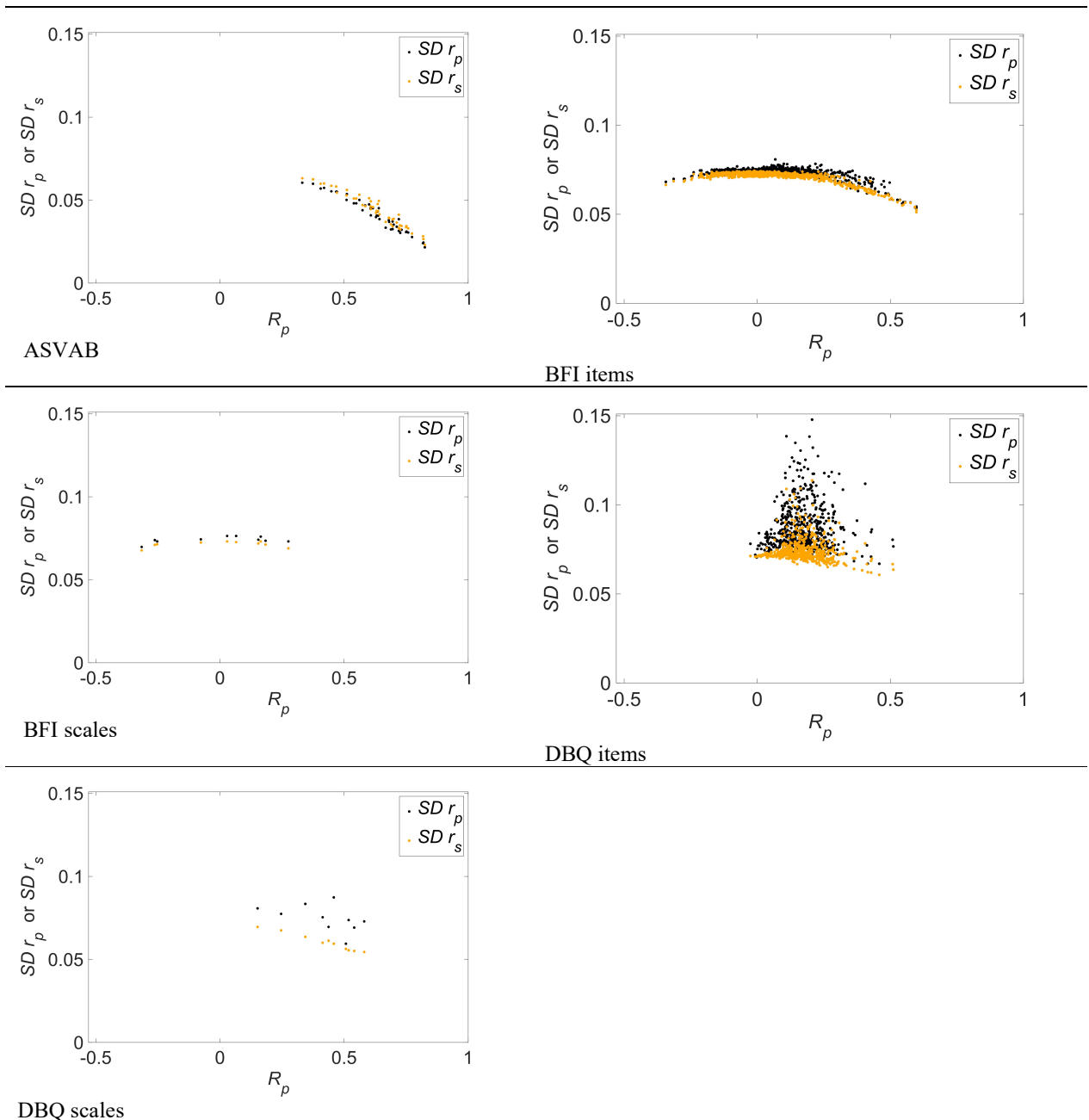

*Figure 7.* Standard deviation (*SD*) of the Pearson correlation coefficient ($r_p$) and the standard deviation of the Spearman correlation coefficient ($r_s$) ($N$ = 200) as a function of the population Pearson correlation coefficient ($R_p$). The population coefficient $R_p$ was defined as the correlation coefficients for the total sample ($N$ = 11,878 for the ASVAB, $N$ = 1,895,753 for the BFI, and $N$ = 9,077 for the DBQ). Top left: Armed Services Vocational Aptitude Battery (ASVAB; 45 correlation coefficients). Top right: Big Five Inventory (BFI) items (946 correlation coefficients). Middle left: BFI scales (10 correlation coefficients). Middle right: Driver Behaviour Questionnaire (DBQ) items (561 correlation coefficients). Bottom left: DBQ scales (10 correlation coefficients).

**Discussion**

The Pearson product-moment correlation coefficient ($r_p$) and the Spearman rank correlation coefficient ($r_s$) are widely used in psychology, with $r_p$ being the most popular. The two coefficients have different goals: $r_p$ is a measure of the degree of linearity between two vectors of data, whereas $r_s$ measures their degree of monotonicity.

The characteristics of $r_p$ and $r_s$ have been widely studied for over 100 years, and in the case of bivariate normality, the distribution of $r_p$ is known exactly (Equation 4). The influence functions of the Pearson and Spearman correlations have been described exactly as well (e.g., Croux & Dehon, 2010). However, several of these features of $r_p$ and $r_s$ may not be known among substantive researchers, and hence our simulations of normally distributed variables are presented as a helpful tutorial. In other words, we illustrated in an intuitive graphical manner the variability, bias, and robustness properties of both coefficients, with a focus on the effect sizes and sample sizes that are likely to occur in psychological research. The relative performance of $r_p$ and $r_s$ in real psychological datasets for different item characteristics, sample sizes, and aggregation methods (i.e., item and scale levels) is intended to facilitate informed decision making regarding when to select $r_p$ and when to select $r_s$.

Our computer simulations showed that for normally distributed variables $r_s$ behaves approximately the same as $r_p$, with $r_s$ being slightly lower and more variable than $r_p$. The difference between the standard deviation of $r_p$ and $r_s$ was minor (< 1%) when the association was weak or moderate in the population ($R_p = 0$ and $R_p = .2$). However, $r_s$ had a substantially higher standard deviation than $r_p$ when the correlation was strong (i.e., a 3 to 4% higher standard deviation when $R_p = .4$) or very strong (i.e., 18% higher standard deviation when $R_p = .8$).

In psychological research, near-normally distributed data, such as the ASVAB test scores, do occur. We showed that for the strongly intercorrelated and approximately normally distributed variables of the ASVAB, $r_p$ slightly outperformed $r_s$ in terms of variability. The expected values of $r_p$ and $r_s$ were almost the same, but the standard deviation of $r_p$ was about 6% lower than the standard deviation of $r_s$. The similarity of $r_s$ and $r_p$ for normally distributed psychometric variables is consistent with empirical sampling research in the physical sciences, where normally distributed variables tend to be common (McDonald & Green, 1960). However, in psychological research, heavy-tailed distributions are common (Blanca et al., 2013; Micceri, 1989). Using a simulation of two correlated variables with heavy-tailed distributions, we showed that $r_s$ was between 13 and 27% less variable than was $r_p$.

The comparative efficacy of $r_p$ versus $r_s$ was further explored in a sampling study of BFI and DBQ survey data at both the item and scale levels. For these survey datasets, $r_s$ turned out to be between 2% and 24% less variable than $r_p$. In fact, for the DBQ dataset, we found that the sample Spearman correlation coefficient ($r_s$) was a more accurate approximation of the population Pearson correlation coefficient ($R_p$) than was the sample Pearson correlation coefficient ($r_p$). This inaccuracy of $r_p$ with respect to $R_p$ was particularly large when the two variables had heavy-tailed distributions (see Figure S7 of supplementary material).

Our simulations further made clear that $r_s$ is robust, while $r_p$ is sensitive to an outlier, even for a sample size as high as 200. Outliers may be caused by a recording error, an error in the experimental procedure, or an accurate representation of a rare case (Cohen et al., 2013). It is likely that real-life data are contaminated with "faulty data" (Spearman, 1910) or an "accidental error" (Spearman, 1904, p. 81), and therefore the robustness of the Spearman estimator ($r_s$) is a virtue for empirical researchers. Using Anscombe's (1960) insurance policy analogy, $r_s$ yields a slight loss of efficiency when bivariate normality assumptions are met, but this seems a small premium given the impressive protection it provides against outliers (Figure 5).

Our study also illustrated the dramatic effect of sample size on the variability of the correlation coefficients. A sample size of 25 yields average errors that are often even larger than the absolute magnitude of the correlation coefficient (e.g., Figure 2; Table S1), which essentially means that the observed correlations are almost meaningless. The standard deviations of $r_s$ and $r_p$ decrease approximately according to the square root of sample size, which means that the standard deviations reduce by approximately 41% when sample size is doubled (cf. Figure S6). In other words, although substantial efficiency gains can be achieved by choosing $r_s$ instead of $r_p$, the effect of sample size is much more dramatic, and therefore we urge researchers to always monitor the confidence interval of their obtained effects.

So, should a practitioner use $r_p$ or $r_s$? Of course, the two correlation coefficients have different goals: $r_p$ represents the strength of the linear relationship between two vectors of data, whereas $r_s$ describes their degree of monotonicity.

Because $r_p$ and $r_s$ have different goals, they strictly ought not to be seen as competing approaches. That is, if one's aim is solely to assess whether the individual sample data points are linearly related (regardless of any nonlinearity that exists), and one's sample size is very large, then $r_p$ should be used. However, it is likely that practitioners are interested in obtaining a high quality correlational measure in terms of low variability, low bias, and high robustness. In such case, $r_s$ clearly has attractive properties compared to $r_p$. If one expects that the two variables have low kurtosis (i.e., normal or platykurtic distributions) and outliers are unlikely to be present, $r_p$ is to be recommended. In other circumstances, $r_s$ seems to be the preferred method because of its superior performance in terms of variability and robustness. The 'embarrassing' failure of $r_p$ to accurately estimate its own population value ($R_p$) in the DBQ dataset, both at the item and at the scale levels, strongly argues in favor of using $r_s$ for heavy-tailed survey data. Note that the behavior of $r_p$ and $r_s$ depends not just on kurtosis, but also on sample size, the population correlation coefficient, and the type of nonlinear relationship between the two variables (see supplementary material). These factors may explain some of the idiosyncratic behaviors of the datasets (see Table 3). Ambiguity arises when having to analyze a large set of variables, whereby half of the data are platykurtic and the other half leptokurtic. In this case, again using Anscombe's (1960) insurance parallel, we recommend using $r_s$ instead of $r_p$, because the premium-protection trade-off is not symmetric. After all, there is a relatively small increase of variability for the variables that are indeed platykurtic, while $r_s$ offers marked robustness to heavy tails and outliers.

There are, of course, a large number of other types of data transformations, such as a logarithmic, multiplicative inverse, or power transformation, that can be successfully applied prior to calculating the Pearson correlation coefficient (Bishara & Hittner, 2012). However, whereas the rank transformation as used in the Spearman correlation coefficient is broadly applicable, other types of data transformation are not. For example, a logarithmic or square root transformation is impossible on negative numbers (unless applying an arbitrary offset), and the multiplicative inverse transformation dilutes any meaningful association when some of the numbers are close to zero. In other words, it is quite possible to mess up one's data by choosing the 'wrong' type of transformation, so that, for example, a normal distribution becomes highly non-normal. As a result, selecting an appropriate nonlinear data transformation requires either prior knowledge of the population distribution or the ethically dubious practice of 'peeking' at the data (Sagarin, Ambler, & Lee, 2014), and it is therefore difficult to come up with systematic meaningful guidelines. In contrast, the Spearman correlation appears to be applicable across a broad array of normal and non-normal distributions.

Alternative measures of association, such as the percentage bend correlation (Wilcox, 1994), the Winsorized correlation (Wilcox, 1993), and the Kendall tau rank correlation coefficient ($r_t$), may be even more robust and efficient than $r_s$ (see Croux & Dehon, 2010). $r_t$ is attractive because it can be interpreted intuitively as the proportion of pairs of observations that are in the same order on both variables minus the proportion that are opposite (Cliff, 1996; Noether, 1981). Other attractive properties of $r_t$ are that it is an unbiased estimator of its population value and that the variance is given in closed form (Esscher, 1924; Fligner & Rust, 1983; Hollander, Wolfe, & Chicken, 2013; Kendall, 1948; Kendall, Kendall, & Babington Smith, 1939; Xu, Hou, Hung, & Zou, 2013). However, Xu et al. (2013) argued that $r_s$ has a lower computational load than $r_t$, and that the variance of $r_s$ can be approximated with high numerical accuracy, leading the authors to conclude that the mathematical advantage of $r_t$ over $r_s$ is not of great importance. Another issue is that $r_t$ converges to markedly different population values than $r_p$ and $r_s$. For typical bivariate normal distributions, $r_p$ and $r_s$ are about 50% greater than $r_t$ (Equations 9 and 10, Fredricks & Nelsen, 2007, see also Figure S16). Because present-day researchers are familiar with interpreting $r_p$ (see Table 1), it seems unlikely that $r_t$ could replace $r_p$. $r_s$ on the other hand has the potential to be used in place of $r_p$, because, as we showed, $r_s$ can surpass $r_p$ in estimating $R_p$. Corrected correlations, such as polychoric correlations, may also be useful alternatives to the Spearman correlation, especially for multivariate applications. Although multivariate methods using the polychoric correlation matrix have been implemented in almost all SEM packages and are still under scrutiny (e.g., Rhemtulla, Brosseau-Liard, & Savalai, 2012; Yuan, Wu, & Bentler, 2011), the polychoric correlation has not yet caught on among substantive researchers (see Table 1).

There are established ways of dealing with outliers, including outlier removal and robust approaches such as least absolute deviation, least trimmed squares, M-estimates, and bounded inference estimators (Cohen et al., 2013; Rousseeuw & Leroy, 2005), or procedures that take into account the structure of the data (Wilcox & Keselman, 2012, see Pernet, Wilcox, & Rousselet, 2012 for an open source MATLAB toolbox). However, removing outliers is an inherently subjective procedure, and retaining too much flexibility could easily lead to inflated effect sizes and false positive inferences (Bakker & Wicherts, 2014; Cohen et al., 2013). It is noted that high kurtosis and outliers can be indicative of problems in the measurement procedure. Subtle changes in questionnaire wording or anchoring can have large effects on the obtained results (Schwarz, 1999). We recommend that researchers remedy the root causes of outliers and high kurtosis before they continue their study.

The choice of correlation coefficient is important not only for establishing bivariate relationships. Psychologists often intend to do follow-up analyses, such as to calculate a percentage of variance explained, to perform an ANOVA or MANOVA, to carry out a meta-analysis of correlation coefficients, or to establish a matrix of correlation coefficients to be submitted to a multivariate statistical method such as principal component analysis, factor analysis, or structural equation modeling. Cliff (1996) argued that perhaps most of the answers that psychologists want to get from their data are ordinal ones, and the data they work with have, at best, ordinal justification. He concluded that ordinal questions should be answered ordinally, instead of trying to answer them with Pearson correlations, mean differences, and parametric techniques. Using ordinal statistics has the added benefit that the inferences remain unchanged if the variables are monotonically transformed (Cliff, 1996). Unfortunately, purely ordinal multivariate statistical methods are rare and generally less developed than traditional parametric methods (for a possible exception using Kendall's tau, see Cliff, 1996).

Indeed, there has been considerable controversy about the use of a rank transformation, because corresponding statistical procedures in complex research designs are sometimes unavailable, inexact, and difficult to interpret (e.g., Fligner, 1981; Sawilowsky, 1990; Zimmerman, 2012). In some cases, the rank transformation may be even entirely inappropriate. For example, when testing the null hypothesis of no interactions in a multifactorial layout, the rank transformation can yield a test statistic that goes to infinity as the sample size increases (Thompson, 1991; see also Akritas, 1993; Sawilowsky, Blair, & Higgins, 1989). Hence, our present results, which favor $r_s$ over $r_p$, seem to lead to a "cul de sac" for researchers in psychology.

However, one could set aside such theoretical constraints, and adopt "a pragmatic sanction" (Stevens, 1951, p. 26). We argue that there is no good reason to stick to $r_p$ for the mere reason that it is consistent with follow-up analyses such as ANOVA and principal component analysis. It is easily forgotten that the assumption of normality is almost always violated in the population, and that calculating $r_p$ on ordinal data, such as those obtained from Likert items, is not strictly permissible anyway (Stevens, 1946). The debate of representational versus pragmatic measurement is a long and bitter one with deep philosophical roots (e.g., Hand, 2004; Michell, 2008; Velleman & Wilkinson, 1993). We support Lord's (1953) pragmatic view that "the numbers don't remember where they came from" (p. 21), and we argue that if $r_s$ outperforms $r_p$ in terms of bias, variability, and robustness, then there is no justifiable reason for not using $r_s$. We illustrate this point by submitting an $r_s$ correlation matrix and an $r_p$ correlation matrix of the DBQ data to a principal component analysis (and see Babakus, Ferguson, & Jöreskog, 1987 and Mittag, 1993, for a similar approach). Results showed that the first six eigenvalues of the $r_p$ correlation matrix were between 26% and 68% more variable than the eigenvalues of the $r_s$ correlation matrix (see Table S3), which means that the factor structure is more stable if researchers simply base their multivariate analyses on the $r_s$ matrix. In some software packages, it is relatively easy to submit the $r_s$ matrix to a multivariate analysis (e.g., in MATLAB *factoran(corr(X, 'type', 'spearman'),2, 'xtype', 'covariance')* performs a maximum likelihood factor analysis on the X matrix, extracting two factors). However, in SPSS, for example, this analysis requires extensive scripting (Garcia-Granero, 2002). Therefore, we recommend the simpler approach of transforming all variables to ranks prior to running the multivariate analysis (e.g., *factoran(tiedrank(X),2)* in the MATLAB command window or *Transform > Rank Cases* from SPSS's pull-down menu). Summarizing, a rank-transformation is an appropriate bridge between non-parametric and parametric statistics (Conover & Iman, 1981).

## Acknowledgements

The datasets used in this research were obtained from the Transport Research Laboratory (2008), the Bureau of Labor Statistics (2002), and the Gosling-Potter Internet Personality Project. The principal investigator of the Gosling-Potter Internet Personality Project can be contacted to access the data from this project (samg@austin.utexas.edu).

Zimmerman, D. W., Zumbo, B. D., & Williams, R. H. (2003). Bias in estimation and hypothesis testing of correlation. *Psicológica: Revista de Metodología y Psicología Experimental, 24*, 133–158.

**Supplementary material**

Table S1

*Means and standard deviations of sample correlation coefficients, and mean absolute difference between sample correlation coefficients and population correlation coefficients (N = 25).*

| | ASVAB | BFI items | BFI scales | DBQ items | DBQ scales |
|---|---|---|---|---|---|
| | **mean across 45 correlations** | **mean across 946 correlations** | **mean across 10 correlations** | **mean across 561 correlations** | **mean across 10 correlations** |
| \|Mean $r_p$\| | .6230 | .1194 | .1739 | .1810 | .4102 |
| \|Mean $r_s$\| | .6104 | .1202 | .1633 | .1749 | .4063 |
| \|Mean $r_p$\|−\|$R_p$\| | −0.0043 | −0.0012 | −0.0039 | 0.0096 | −0.0096 |
| \|Mean $r_s$\|−\|$R_s$\| | −0.0177 | −0.0020 | −0.0057 | 0.0126 | −0.0094 |
| $SD\ r_p$ | .1214 | .2094 | .2122 | .2315 | .1943 |
| $SD\ r_s$ | .1309 | .2057 | .2053 | .2125 | .1756 |
| Mean \|$r_p$−$R_p$\| | .0954 | .1689 | .1709 | .1895 | .1561 |
| Mean \|$r_p$−$R_s$\| | .0962 | .1690 | .1713 | .1898 | .1572 |
| Mean \|$r_s$−$R_p$\| | .1038 | .1655 | .1652 | .1739 | .1414 |
| Mean \|$r_s$−$R_s$\| | .1031 | .1653 | .1649 | .1742 | .1401 |

*Note.* $r_p$ = sample Pearson correlation coefficient, $R_p$ = population Pearson correlation coefficient, $r_s$ = sample Spearman correlation coefficient, $R_s$ = population Spearman correlation coefficient, ASVAB = Armed Services Vocational Aptitude Battery, BFI = Big Five Inventory, DBQ = Driver Behaviour Questionnaire. The absolute means, standard deviations, and mean absolute differences were calculated for each off-diagonal item of the correlation matrix (45, 946, 10, 561, & 10 correlations for the ASVAB, BFI items, BFI scales, DBQ items, & DBQ scales, respectively) and subsequently averaged. $R_p$ and $R_s$ were defined as the correlation coefficients for the total sample ($N$ = 11,878 for the ASVAB, $N$ = 1,895,753 for the BFI, & $N$ = 9,077 for the DBQ). The results were based on 50,000 samples of $N$ = 25. When the correlation matrix could not be calculated due to the small sample size, the sampling was repeated.

Table S2

*Means and standard deviations of sample correlation coefficients, and mean absolute difference between sample correlation coefficients and population correlation coefficients (N = 1,000).*

| | ASVAB | BFI items | BFI scales | DBQ items | DBQ scales |
|---|---|---|---|---|---|
| | **mean across 45 correlations** | **mean across 946 correlations** | **mean across 10 correlations** | **mean across 561 correlations** | **mean across 10 correlations** |
| \|Mean $r_p$\| | .6273 | .1206 | .1778 | .1709 | .4193 |
| \|Mean $r_s$\| | .6277 | .1222 | .1690 | .1621 | .4154 |
| \|Mean $r_p$\|−\|$R_p$\| | .0000 | .0000 | .0000 | −0.0004 | −0.0005 |
| \|Mean $r_s$\|−\|$R_s$\| | −0.0004 | 0.0000 | 0.0000 | −0.0001 | −0.0003 |
| $SD\ r_p$ | .0182 | .0327 | .0332 | .0401 | .0348 |
| $SD\ r_s$ | .0193 | .0319 | .0319 | .0331 | .0269 |
| Mean \|$r_p$−$R_p$\| | .0145 | .0261 | .0265 | .0320 | .0277 |
| Mean \|$r_p$−$R_s$\| | .0195 | .0267 | .0284 | .0338 | .0342 |
| Mean \|$r_s$−$R_p$\| | .0201 | .0261 | .0274 | .0291 | .0294 |
| Mean \|$r_s$−$R_s$\| | .0154 | .0255 | .0255 | .0264 | .0214 |

*Note.* $r_p$ = sample Pearson correlation coefficient, $R_p$ = population Pearson correlation coefficient, $r_s$ = sample Spearman correlation coefficient, $R_s$ = population Spearman correlation coefficient, ASVAB = Armed Services Vocational Aptitude Battery, BFI = Big Five Inventory, DBQ = Driver Behaviour Questionnaire. The absolute means, standard deviations, and mean absolute differences were calculated for each off-diagonal item of the correlation matrix (45, 946, 10, 561, & 10 correlations for the ASVAB, BFI items, BFI scales, DBQ items, & DBQ scales, respectively) and subsequently averaged. $R_p$ and $R_s$ were defined as the correlation coefficients for the total sample ($N$ = 11,878 for the ASVAB, $N$ = 1,895,753 for the BFI, & $N$ = 9,077 for the DBQ). The results were based on 50,000 samples of $N$ = 1,000.

Table S3

*Means and standard deviations of the first six eigenvalues of the 34 x 34 correlation matrices of the Driver Behaviour Questionnaire (DBQ).*

| | $r_p$ matrices Mean (*SD*) | $r_s$ matrices Mean (*SD*) | $R_p$ matrix | $R_s$ matrix |
|---|---|---|---|---|
| Eigenvalue 1 | 6.977 (0.896) | 6.666 (0.577) | 6.830 | 6.547 |
| Eigenvalue 2 | 2.910 (0.317) | 2.704 (0.234) | 2.673 | 2.517 |
| Eigenvalue 3 | 1.845 (0.155) | 1.689 (0.092) | 1.274 | 1.238 |
| Eigenvalue 4 | 1.606 (0.092) | 1.530 (0.069) | 1.256 | 1.205 |
| Eigenvalue 5 | 1.459 (0.073) | 1.416 (0.058) | 1.206 | 1.174 |
| Eigenvalue 6 | 1.346 (0.066) | 1.321 (0.052) | 1.049 | 1.061 |

*Note.* $r_p$ = sample Pearson correlation coefficient, $R_p$ = population Pearson correlation coefficient, $r_s$ = sample Spearman correlation coefficient, $R_s$ = population Spearman correlation coefficient. The sample correlation coefficients were based on 50,000 samples of $N = 200$. $R_p$ and $R_s$ were defined as the correlation coefficients for the total sample ($N = 9,077$).

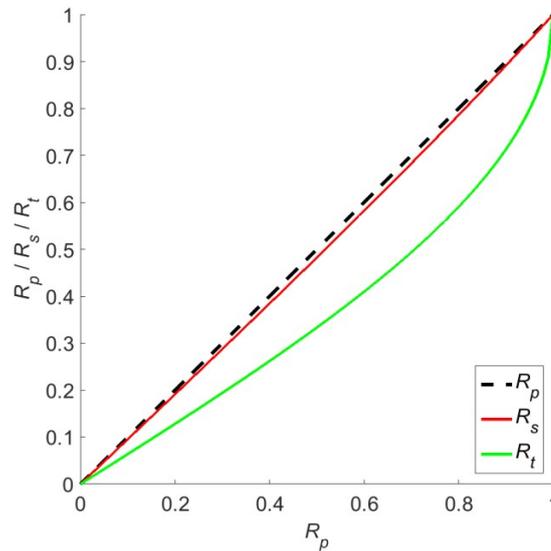

*Figure S1*. The red line is the relationship between the population Spearman correlation coefficient ($R_s$) and the population Pearson correlation coefficient ($R_p$) in the case of bivariate normality. The green line is the relationship between the population Kendall's tau ($R_t$) and $R_p$ in the case of bivariate normality. The dashed black line represents $R_p$ versus $R_p$ and therefore runs diagonally.

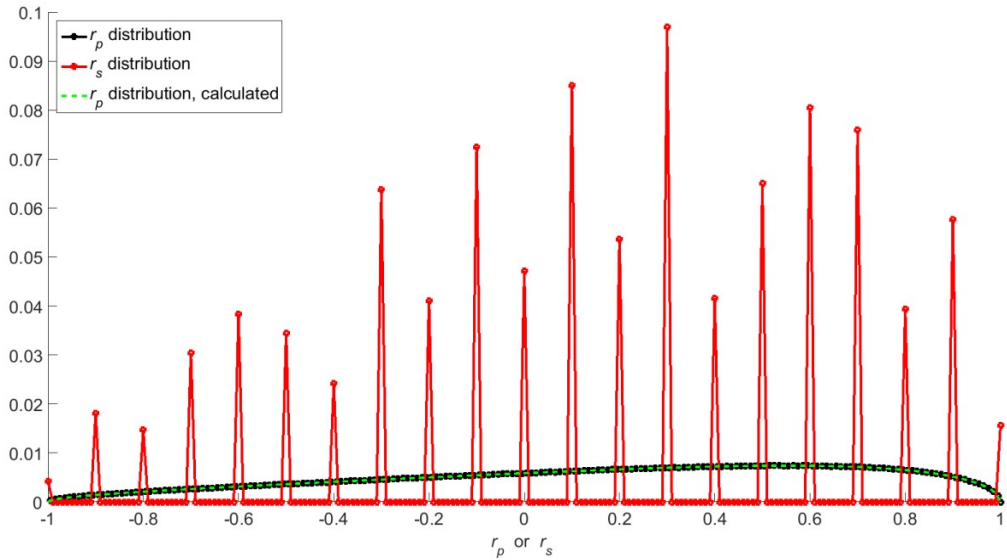

*Figure S2.* Simulation results for normally distributed variables having a population Pearson correlation coefficient of .2 ($R_p$ = .2). The figure shows the distribution of the Pearson correlation coefficient ($r_p$) and the Spearman correlation coefficient ($r_s$) for a sample size ($N$) of 5. The distribution was obtained from a simulation of $10^7$ repetitions. The resolution of the distribution was 0.01. The results have been normalized so that the sum of the 201 counts equaled 1. The figure also depicts the exact distribution of $r_p$ calculated with Equation 4, which lies almost exactly on top of the results of the simulation study.

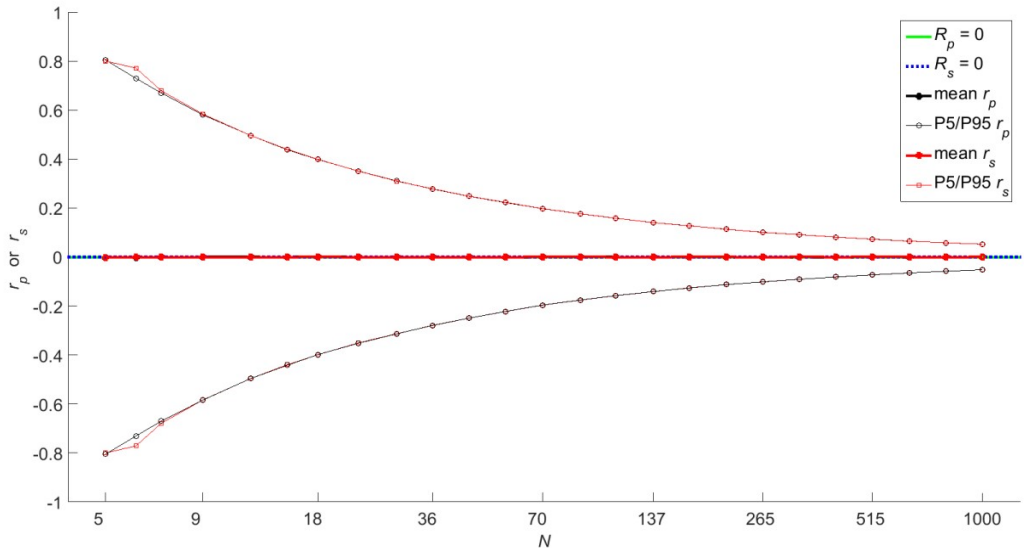

*Figure S3.* Simulation results for normally distributed variables having a population Pearson/Spearman correlation coefficient of 0 ($R_p = R_s = 0$). The figure shows the mean, 5th percentile (P5), and 95th percentile (P95) of the Pearson correlation coefficient ($r_p$) and the Spearman correlation coefficient ($r_s$) as a function of sample size ($N$).

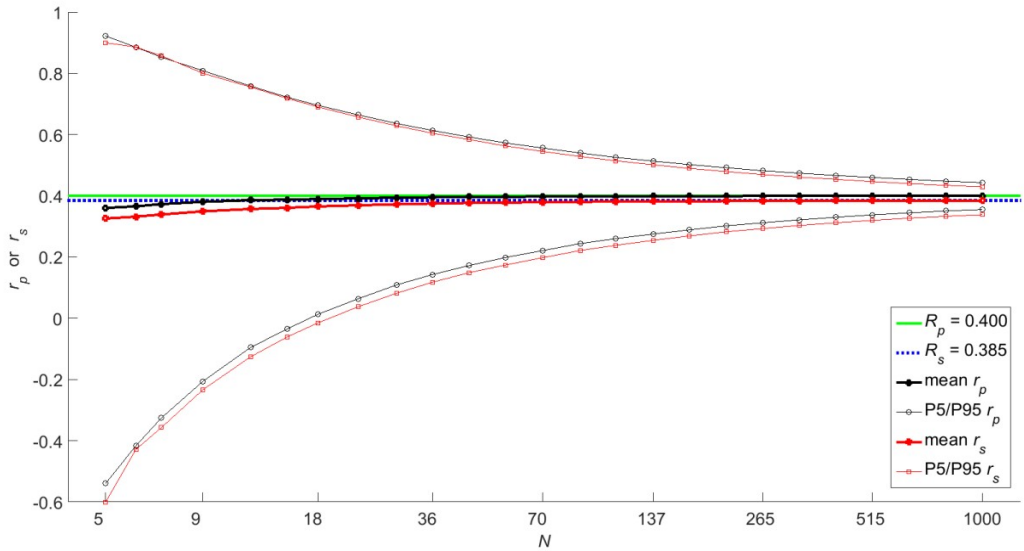

*Figure S4.* Simulation results for normally distributed variables having a population Pearson correlation coefficient of .4 ($R_p$ = .4). The population Spearman correlation coefficient ($R_s$) was calculated according to Equation 9. The figure shows the mean, 5th percentile (P5), and 95th percentile (P95) of the Pearson correlation coefficient ($r_p$) and the Spearman correlation coefficient ($r_s$) as a function of sample size ($N$).

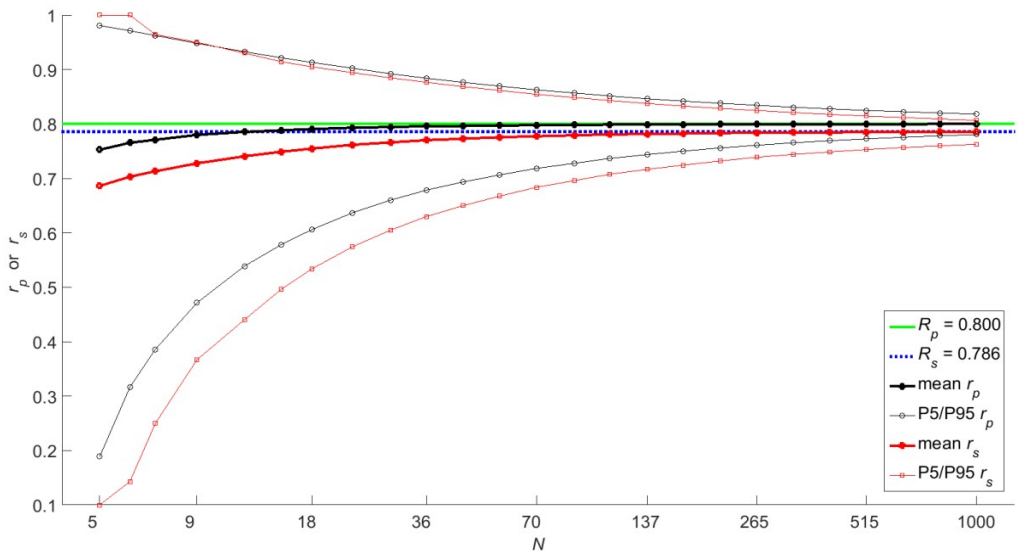

*Figure S5.* Simulation results for normally distributed variables having a population Pearson correlation coefficient of .8 ($R_p$ = .8). The population Spearman correlation coefficient ($R_s$) was calculated according to Equation 9. The figure shows the mean, 5th percentile (P5), and 95th percentile (P95) of the Pearson correlation coefficient ($r_p$) and the Spearman correlation coefficient ($r_s$) as a function of sample size ($N$).

**Supplementary material explaining the behavior of $r_p$ and $r_s$ for non-normal distributions and non-linear associations**

We explored the behavior of $r_p$ and $r_s$ for two correlated variables ($R_p = .4$) having a $\chi^2$ distribution. Figure S6 shows the standard deviation of $r_p$ as a function of sample size. It can be seen that the lower the degrees of freedom of the $\chi^2$ distributions (and hence the greater the skewness and kurtosis of the two variables), the more variable $r_p$ is. A $\chi^2$ distribution with 32 degrees of freedom closely resembles a normal distribution, which is why the standard deviation of $r_p$ and $r_s$ are almost the same in that case.

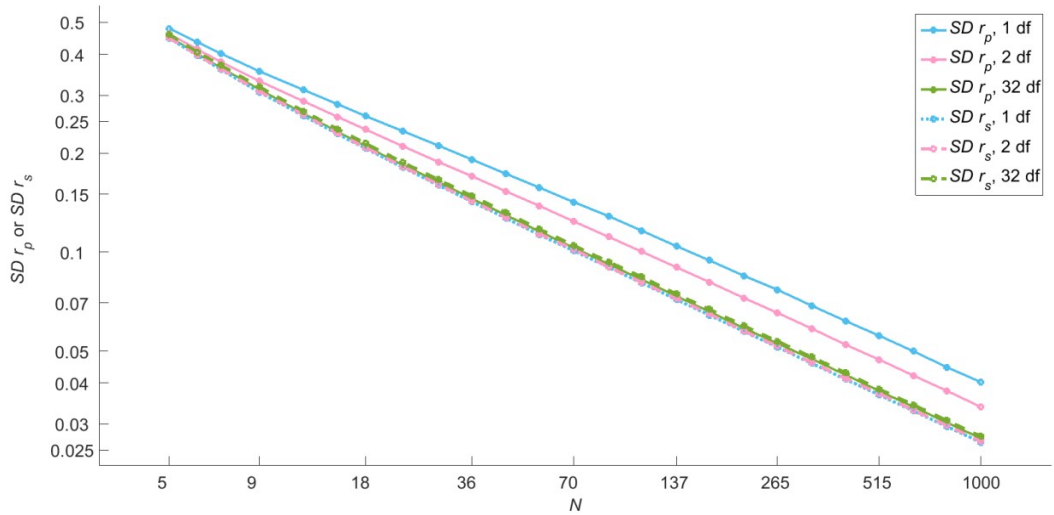

*Figure S6*. Standard deviation of $r_p$ and standard deviation of $r_s$ for three approximated $\chi^2$ distributions with different degrees of freedom (*df*) and population Pearson correlation coefficient of .4. The population skewness is 2.83, 2.00, and 0.50 for 1 *df*, *2f*, and 32 *df*, respectively. The population kurtosis is 15, 9, and 3.38 for 1 *df*, 2 *df*, and 32 *df*, respectively. A $\chi^2$ distribution with 2 *df* is an exponential distribution (see also Figures 3 & 4). The distributions were created using a method by Headrick (2002).

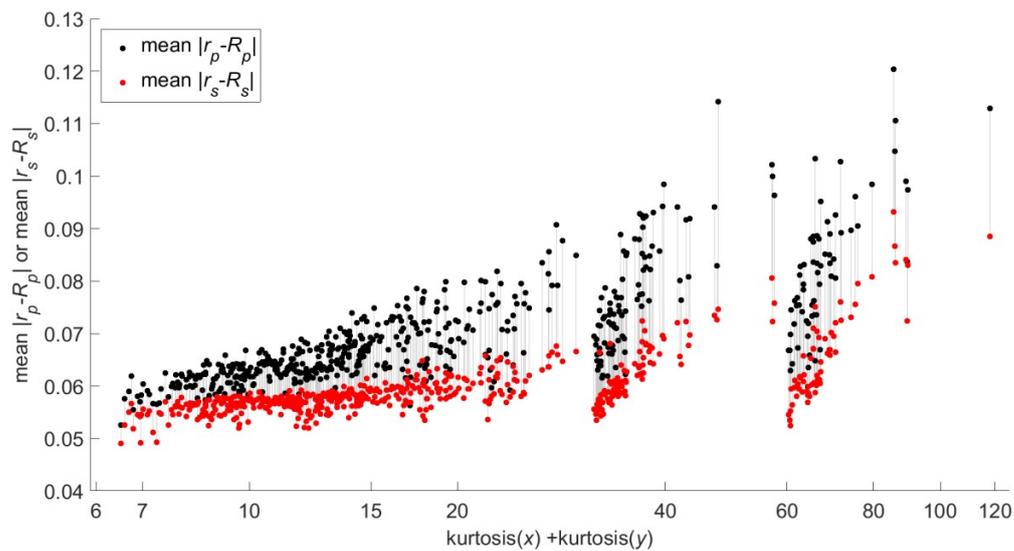

*Figure S7.* Mean absolute differences between sample Pearson correlation coefficient ($r_p$) and population Pearson correlation coefficient ($R_p$) (black dots) and mean absolute differences between sample Spearman correlation coefficient ($r_s$) and population Spearman correlation coefficient ($R_s$) (red dots) for pairs of variables ($x$, $y$) of the Driver Behaviour Questionnaire (DBQ) dataset as a function of the population kurtosis of $x$ plus the population kurtosis of $y$ ($N = 9{,}077$). The mean absolute differences for $r_p$ and $r_s$ are connected by a gray vertical line for each pair of variables. These results were based on 50,000 pairs of samples of $N = 200$. The $x$-axis is logarithmic.

When the two variables have a joint normal distribution, then the expected value of $y$ for a given $x$ is a linearly related to $x$ (e.g., Bertsekas & Tsitsiklis, 2014). Figure S8 below illustrates an $R_p$ of .95 and an $R_p$ of .50 for two variables having a mean of 0 and a standard deviation of 1. The cyan (corresponding to $R = .95$) and yellow (corresponding to $R_p = .5$) lines represent the means for a given $x$. The slopes of the lines are equal to the correlation coefficient.

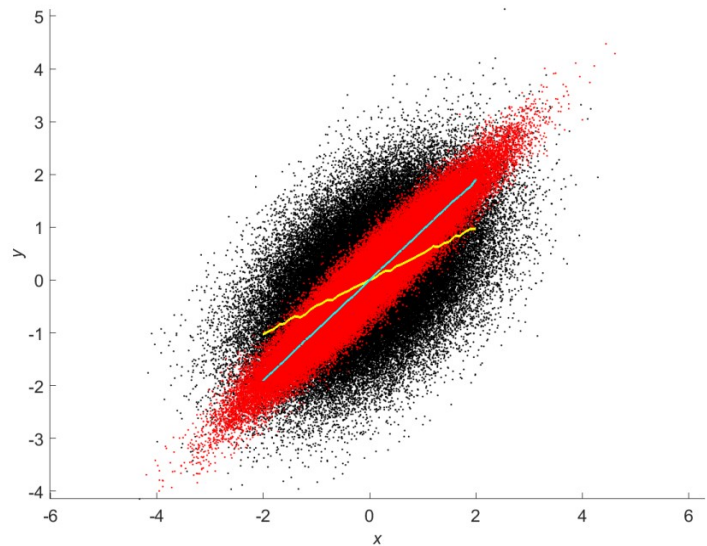

*Figure S8.* Simulation of two normally distributed variables (*x* & *y*) having a population Pearson correlation coefficient $R_p$ = .95 (visualized for *N* = 100,000) and $R_p$ = .5 (visualized for *N* = 100,000). The lines represent the mean value of *y* for a given *x* for $R_p$ = .95 and $R_p$ = .5. The means are calculated for bins of *x* that are 0.1 wide.

A nonlinear relationship between two variables can only occur when at least one of the two variables is non-normally distributed. However, even two highly skewed variables can be linearly related, if they have the same type of distribution. Figure S9 illustrates the relationship between two items of the Driver Behaviour Questionnaire (DBQ). Both variables are skewed (skewness of *x* = 2.85; skewness of *y* = 1.45; skewness of *x* = 14.5; skewness of *y* = 4.83), yet their relationship is approximately linear. In other words, when two variables are normally distributed, then their relationship is linear. But two highly skewed distributions are not necessarily non-linearly related.

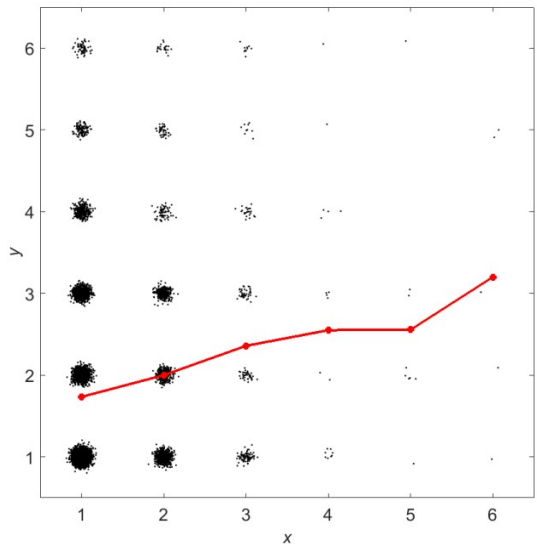

*Figure S9.* Relationship between two items of the Driver Behaviour Questionnaire (*x* represents the response to the item "Brake too quickly on a slippery road, or steer the wrong way into a skid"; *y* represents the response to the item "Forget where you left your car in a car park"). Noise with a random distribution and a standard deviation of 0.05 is added for each response to prevent overlap of dots ($R_p$ = .128; $R_s$ = .119; *N* = 9,077). The line represents the mean value of *y* for a given *x*.

We also carried out simulations to explore the effect of the population correlation coefficient ($R_p$). Figures S10 and S12 illustrate the relationship that we generated, with *x* and *y* being exponentially distributed and $R_p$ = .2 and $R_p$ = .8, respectively. The simulation results are provided in Figures S11 and S13, respectively.

It is possible to devise nonlinear relationships where $r_p$ is considerably less variable than $r_s$. Figures S14 and S15 show results for a nonlinear relationship where an exponential distribution is combined with a beta distribution having a negative skewness (−0.85). So, variables having high skewness or high kurtosis can still yield a stable $r_p$.

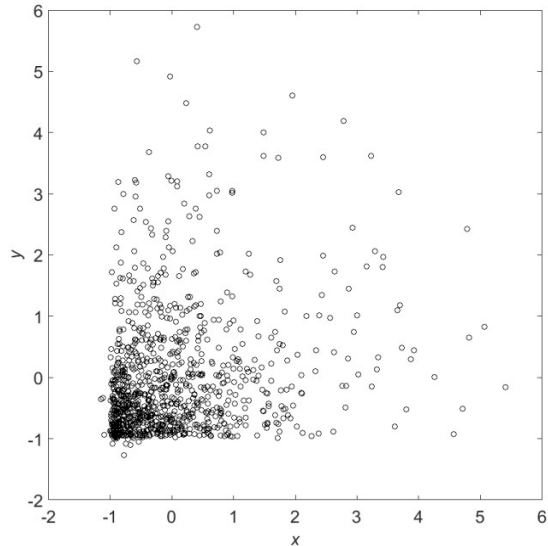

*Figure S10.* Depiction (using $N = 1,000$) of two correlated variables having an exponential distribution with population Pearson correlation coefficient ($R_p$) of .2. $R_p$ was obtained by calculating $r_p$ for a sample of $N = 10^7$ pairs.

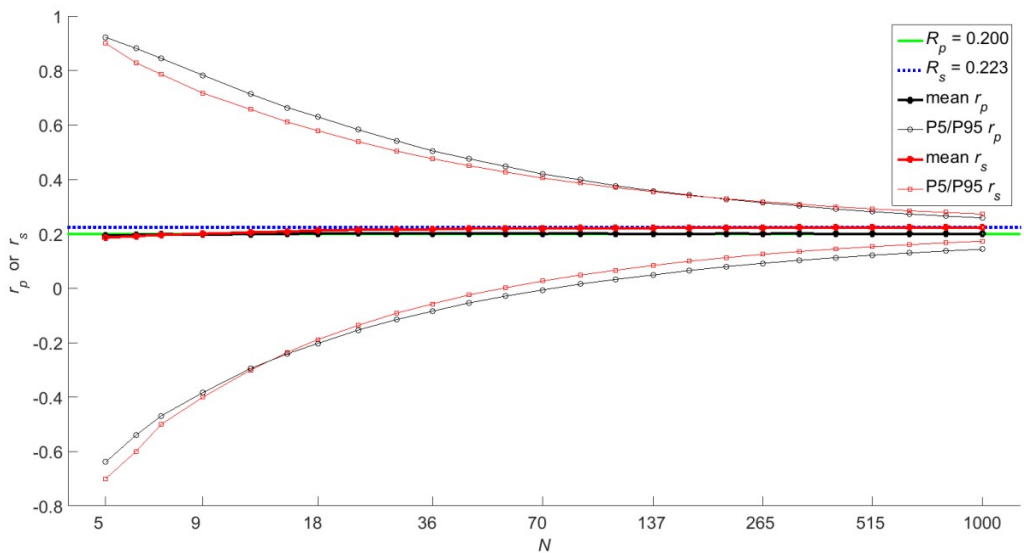

*Figure S11.* Simulation results for two correlated variables having an exponential distribution (see Figure S10 for a large-sample illustration of the distribution). The figure shows the mean, 5th percentile (P5), and 95th percentile (P95) of the Pearson correlation coefficient ($r_p$) and the Spearman correlation coefficient ($r_s$) as a function of sample size ($N$). The population coefficients $R_p$ and $R_s$ were obtained by calculating $r_p$ and $r_s$, respectively, for a sample of $N = 10^7$ pairs.

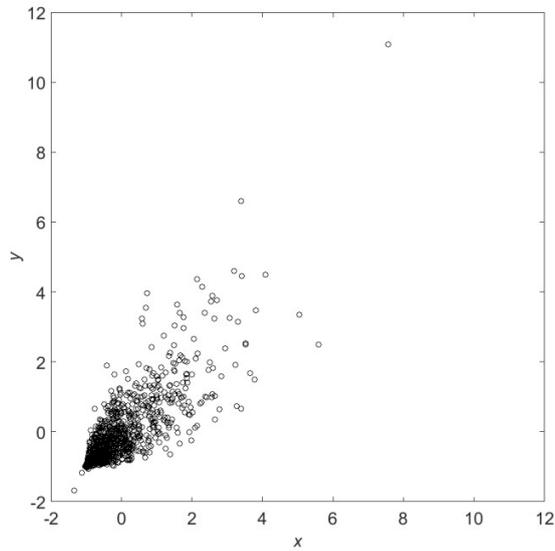

*Figure S12.* Depiction (using $N$ = 1,000) of two correlated variables having an exponential distribution with population Pearson correlation coefficient ($R_p$) of .8. $R_p$ was obtained by calculating $r_p$ for a sample of $N = 10^7$ pairs.

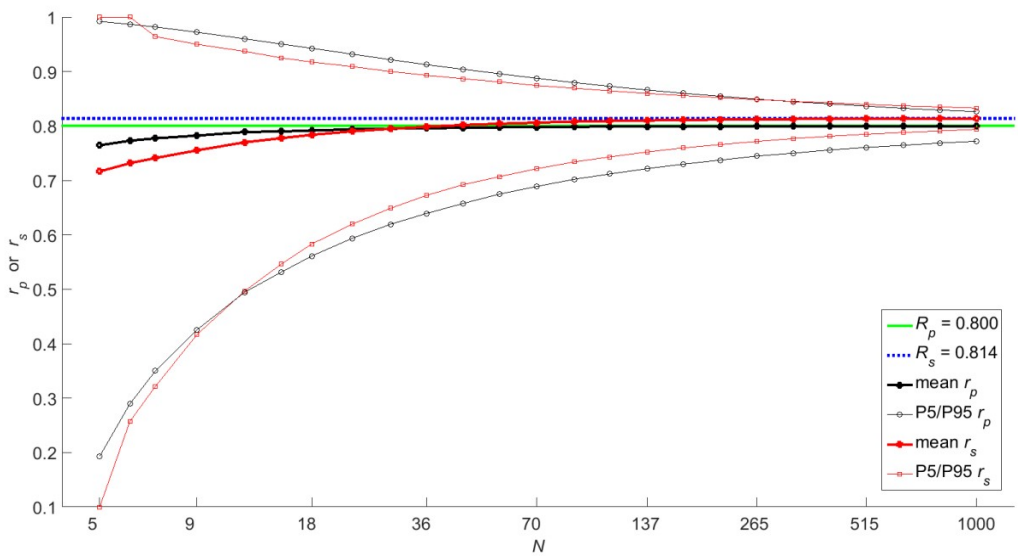

*Figure S13.* Simulation results for two correlated variables having an exponential distribution (see Figure S12 for a large-sample illustration of the distribution). The figure shows the mean, 5th percentile (P5), and 95th percentile (P95) of the Pearson correlation coefficient ($r_p$) and the Spearman correlation coefficient ($r_s$) as a function of sample size ($N$). The population coefficients $R_p$ and $R_s$ were obtained by calculating $r_p$ and $r_s$, respectively, for a sample of $N = 10^7$ pairs.

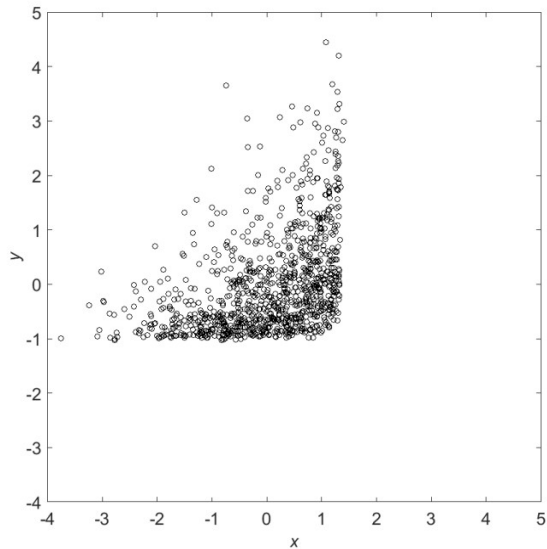

*Figure S14.* Depiction (using $N = 1,000$) of a nonlinear relationship between two variables. The variable $x$ has a population skewness of $-0.85$ and a population kurtosis of 3.22, whereas the variable $y$ has a population skewness of 2 and a population kurtosis of 9 ($R_p = .417$). These population coefficients were calculated for a sample of $N = 10^7$ pairs.

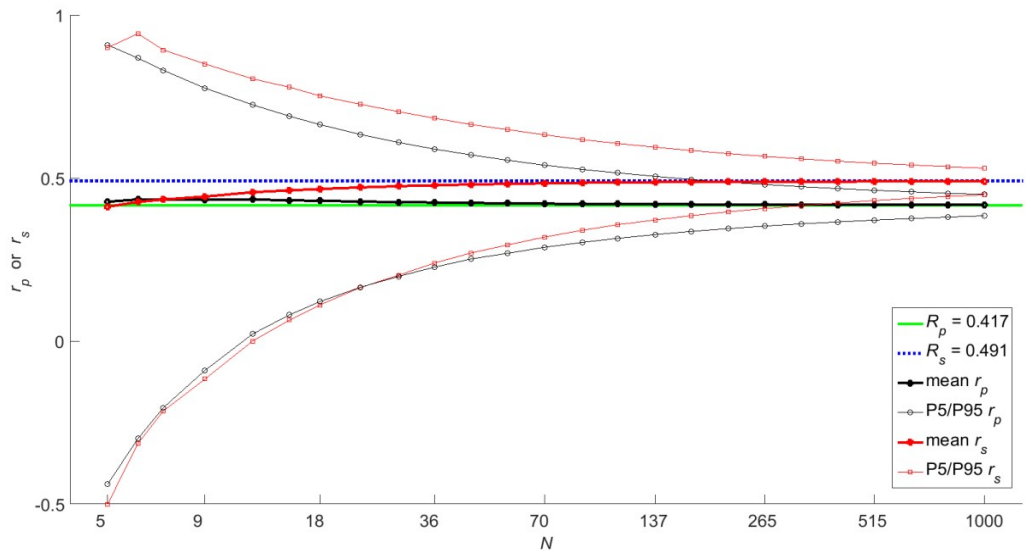

*Figure S15.* Simulation results for a nonlinear relationship between two variables (see Figure S14 for a large-sample illustration of the distribution). The figure shows the mean, 5th percentile (P5), and 95th percentile (P95) of the Pearson correlation coefficient ($r_p$) and the Spearman correlation coefficient ($r_s$) as a function of sample size ($N$). The population coefficients $R_p$ and $R_s$ were obtained by calculating $r_p$ and $r_s$, respectively, for a sample of $N = 10^7$ pairs.

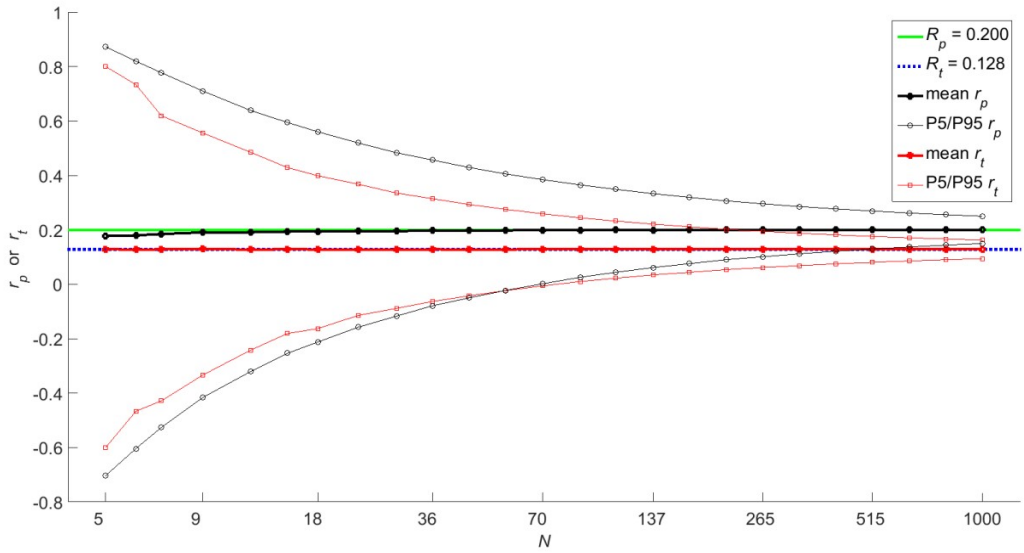

*Figure S16.* Simulation results for normally distributed variables having a population Pearson correlation coefficient of .2 ($R_p$ = .2). The figure shows the mean, 5th percentile (P5), and 95th percentile (P95) of the Pearson correlation coefficient ($r_p$) and the Kendall tau rank correlation coefficient ($r_t$) as a function of sample size ($N$). The population Kendall correlation coefficient ($R_t$) was calculated according to Equation 10.

MATLAB code for producing the figures in this article can be found here
https://supp.apa.org/psycarticles/supplemental/met0000079/met0000079_supp.html